\documentclass[letterpaper, 12 pt, draftcls, onecolumn]{ieeetran}
\IEEEoverridecommandlockouts
 \usepackage{tabularx}
 \usepackage{lscape}
 \usepackage{rotating}
 \usepackage{epstopdf}
 \usepackage{graphicx}
 \usepackage[centertags]{amsmath}
 \usepackage{amsfonts}
 \usepackage{amssymb}
\usepackage{subfigure}
 \usepackage{amsthm}
 \usepackage{epstopdf}
 \usepackage{epsfig}
 \usepackage{graphicx}
 \usepackage[centertags]{amsmath}
 \usepackage{amsfonts}
 \usepackage{amssymb}
 \usepackage{cite}
 \usepackage{multirow}
 \usepackage{amsmath}
 \usepackage{amsthm}
 \usepackage[table]{xcolor}
 \usepackage{float}
 \usepackage{amsfonts,amssymb,amsmath,latexsym}
 \usepackage{acronym}
 \usepackage{enumerate}
 \usepackage{comment}
 \usepackage{epsfig}
 \usepackage{subfigure}
 
\newtheorem{theorem}{Theorem}

\newtheorem{remark}{Remark}

\title{\LARGE \bf
Particle Filter-Based Fault Diagnosis of Nonlinear Systems Using a Dual
Particle Filter Scheme*}
\author{Najmeh Daroogheh$^{1}$,  Nader Meskin$^{2}$ and Khashayar Khorasani$^{1}$
\thanks{*This publication was made possible by NPRP grant No. 4-195-2-065 from the
Qatar National Research Fund (a member of Qatar Foundation). The
statements made herein are solely the responsibility of the authors.}
\thanks{$^{1}$N. Daroogheh and K. Khorasani are with the Department  of Electrical and Computer  Engineering, Concordia University, Montreal, Quebec, Canada
        {\tt\small n\_daroog@encs.concordia.ca and kash@ece.concordia.ca}}%
\thanks{$^{2}$N. Meskin is with the Department of Electrical Engineering, Qatar University,
        Doha, Qatar
        {\tt\small nader.meskin@qu.edu.qa}}%
}

\renewcommand{\baselinestretch}{1.15}

\begin{document}
\maketitle
\begin{abstract}
In this paper, a dual estimation methodology is developed for both time-varying parameters and states of a nonlinear stochastic system based on the Particle Filtering (PF) scheme. {Our developed methodology is based on} a concurrent implementation of state and parameter estimation filters as opposed to using a single filter for simultaneously estimating the augmented states and parameters. The convergence and stability of our proposed dual estimation strategy are shown formally to be guaranteed under certain conditions. The ability of our developed dual estimation method is testified to handle simultaneously and efficiently the states and time-varying parameters of a nonlinear system in a context of health monitoring which employs a unified approach to fault detection, isolation and identification is a single algorithm. The performance capabilities of our proposed fault diagnosis methodology is demonstrated and evaluated by its application to a gas turbine engine through accomplishing state and parameter estimation under simultaneous and concurrent component fault scenarios. Extensive simulation results are provided to substantiate and justify the superiority of our proposed fault diagnosis methodology when compared with another well-known alternative diagnostic technique that is available in the literature.
\end{abstract}
\section{INTRODUCTION}
Systems state estimation is a fundamental problem in control, signal processing, and fault diagnosis fields \cite{state1}. Investigations on both linear and nonlinear state estimation and filtering in stochastic environments have been an active area of research during the past several decades. Linear state estimation methods use a simpler representation of an actual nonlinear system and can provide an acceptable performance only locally around an operating point and in the steady state operational condition of the system. However, as nonlinearities in the system dynamics become dominant, the performance of linear approaches deteriorates and linear algorithms will not necessarily converge to an accurate solution. Although an optimal state estimation solution for linear filtering methods exists, nonlinear filtering methods suffer from generating sub-optimal or near-optimal solutions. Consequently, investigation of nonlinear estimation and filtering problems remain a challenging research area.

Numerous studies have been conducted in the literature to solve and analyze standard nonlinear filtering problems \cite{MS_H_25,MS_H_2,Orchard_2,MS_H_5,MS_H_23,MS_H,cubature}. These methods can be broadly categorized into \cite{MS_H}: (a) linearization methods (extended Kalman filters (EKF)) \cite{MS_H_25}, (b) approximation methods using finite-dimensional nonlinear filters \cite{MS_H_2}, (c) particle filter (PF) methods as one of the most popular Bayesian recursive methods for state estimation \cite{Orchard_2}, (d) classical partial differential equation (PDE) methods for approximating a solution to the Zakai equation  \cite{MS_H_5}, (e) Wiener chaos expansion methods \cite{MS_H_23}, (f) moment methods \cite{MS_H}, and (g) high dimensional nonlinear Kalman filter methods known as the Cubature Kalman filters \cite{cubature}.

One of the most important recent applications of nonlinear filtering methods is in the area of fault diagnosis of dynamical systems that can include fault detection, isolation, and identification (FDII) modules. Diagnosis methods that are based on linearization techniques suffer from poor detection and high rates of false alarms. Therefore, Monte Carlo filtering approach based on particle filters was first proposed in \cite{fault_7} to address the fault detection and isolation problem in nonlinear systems. In this work, the negative log-likelihood, which is calculated for a predefined time window, is considered as a measure for the fault detection. The fault isolation was achieved by using the augmentation of the fault parameters vector to the system states to perform the estimation task. However, the augmented state space model tends to increase the dimensionality of the model and as a result increases the number of required particles for achieving a sufficiently accurate result. For decreasing the computational burden of this method, the augmented model is used only after the fault detection stage and for only the fault isolation stage. An external covariance adjustment loop was added to this augmented model in \cite{Orchard_Thesis} to enable the estimation algorithm to track changes in the system parameters in case of fault occurrences.

The combination of a particle filtering algorithm and the log-likelihood ratio (LLR) test in the multiple model environment, has led to the development of sensor/actuator FDI scheme in \cite{fault_8} for a general class of nonlinear non-Gaussian dynamical systems but with the assumption of full state measurements. The fault detection problem recently is addressed for a mobile robot based on the combination of the negative LLR test and particle filtering approach in \cite{fault_3}. However, both of the methods in \cite{fault_8}, and \cite{fault_3} suffer from the high computational burden for on-line implementation of the algorithms. Hence, the idea of parallelized particle filters for on-line fault diagnosis is introduced in \cite{fault_3} to improve the performance of the algorithm.

A PF-based robust navigation approach was proposed in \cite{fault_10} to address multiple and simultaneous faults occurrences in both actuators and sensors in an underwater robot where an anomaly is modeled by a switching-mode hidden Markov system. The component and actuator fault detection and isolation of a point mass satellite was tackled in \cite{fault_5} by introducing several particle filters that run in parallel and each  rejects a different subset of the faults.

Generally, the main issues with applying standard particle filters to the problem of fault diagnosis problem can be stated as follows \cite{fault_9}: (i) False diagnosis decisions due to low probabilities of transitions to fault states when there are fewer samples of states, and (ii) The exponential growth of the required samples for accurately approximating the {\it{a posteriori}} distributions as dimensionality of the estimation problem increases. The risk-sensitive PF is introduced to address the first problem and the variable resolution PF is developed to overcome the second problem in \cite{fault_4}. Moreover, the Gaussian PF (GPF) is also introduced in \cite{fault_12} as an efficient algorithm for performing fault diagnosis of hybrid systems faster than traditional methods that are based on PFs. Finally, the sample impoverishment problem in particle filters due to fault occurrence in a hybrid system is addressed in \cite{fault_2}. The developed algorithm enables the PF method to be implemented by fewer number of particles even under faulty conditions.

{In this work, our objective is to address the above problems by specifically utilizing particle filters. This is accomplished through development of a PF-based dual state/parameter estimation scheme for the system component fault diagnosis. The developed dual estimation scheme relies on an on-line estimation of the system states as well as the system health parameters. These parameters are not necessarily fixed and their time variations are governed and influenced by the fault vector that directly affects them.}

 The on-line estimation of the system time-varying parameters by using particle filters is a challenging and active area of research. There are two main classes of PF-based parameter estimation algorithms (for on-line as well as off-line implementations) \cite{doucet_2003} known as Bayesian and maximum likelihood (ML) approaches. In the Bayesian approach, a prior distribution is considered for the unknown parameters and the posterior distribution of the parameters is approximated given the observations \cite{west,west_2011}, whereas in the ML approach the estimated parameter is the maximizing argument of the likelihood function given the observations \cite{Doucet, poyi_2005, poyi_2011, EM_shon}. In the ML framework for parameter estimation, the maximization of any cost function can be performed based on gradient-based search methods \cite{Doucet}. On the other hand, expectation maximization (EM) methods are only applicable for maximization of the likelihood functions \cite{EM_shon}. However, EM methods are not suitable for on-line applications due to their high computational cost for implementation. The recursive maximum likelihood method (RML) is recognized as a promising method for on-line parameter estimation based on a stochastic gradient algorithm \cite{poyi_2005}. In order to avoid the direct computation of the likelihood function gradient, an alternative method is proposed in \cite{poyi_gradientfree} that is known as the gradient-free ML parameter estimation. Despite the above, the on-line ML methods suffer from the practical point of view of slow convergence rates and requiring large number of particles to achieve accurate estimates \cite{olsson}.

In the Bayesian framework, on-line implementation of particle filter-based parameter estimation algorithms are computationally intensive \cite{doucet2012}. A general method that is capable of {\it{simultaneously}} estimating the static (i.e. , constant or fixed) parameters and time-varying states of a system is developed in \cite{Parameter}. The work is based on the sequential Monte Carlo (SMC) method in which an artificial dynamic evolution model is considered for the unknown model parameters. In order to overcome the degeneracy concerns arising from the particle filtering, kernel smoothing technique as a method for smoothing the approximation of the parameters conditional density has been utilized in \cite{west}. The estimation algorithm is further improved by re-interpretation of the artificial evolution algorithm according to the shrinkage scaling concept. However, the proposed method in \cite{west} is only applicable for estimating {\it{fixed parameters}} of the system and it uses the augmented state/parameter vector for the estimation task.

{Our main goal in this work is to extend the Bayesian framework that is proposed in \cite{west} for both state and parameter estimation by invoking our proposed modified artificial evolution law that is inspired from the  prediction error (PE) concept. This modification enables the scheme to track and estimate the {\it{time-varying}} parameters of the system. The PE is used for both off-line and on-line system identification by incorporating a quadratically convergent scheme \cite{Walid,PEM}. The concept of the kernel mean shrinkage (KS) is also used in our proposed scheme to avoid over-dispersion in the variance of the estimated parameters.}

Component fault diagnosis of dynamical systems can be achieved through dual state/parameter estimation methods \cite{isermann_fault, ehsan} by first detecting the faulty competent and then by determining its severity and isolating its location. In this work, our proposed dual state and parameter estimation strategy is applied for performing component fault diagnosis of a gas turbine engine \cite{esmaeil}. This is accomplished by utilizing a multiple-model approach as developed in \cite{asme, esmaeil}.
{Although the multiple-model based approach can detect and isolate faults based on a set of predefined faulty models, the utilization of parameter estimation schemes embedded in our proposed dual state/parameter estimation methodology enables the diagnostic scheme to \textit{not only} detect and isolate the faults \textit{but also} accurately identify the type and severity of simultaneous fault scenarios in the gas turbine system. The simultaneous detection, isolation, and identification of faults are the main feature that distinguishes the current work with the works conducted in \cite{asme, esmaeil}. Moreover, the main methodology that is developed in these references is based on Kalman filters as opposed to the particle filtering methodology that is developed on the current work.}

Based on the above discussion, the \underline{main contributions} of this paper can be summarized as that of utilizing nonlinear Bayesian and Sequential Monte Carlo (SMC) methods to develop, design, analyze, and implement a unified framework for both states and parameters estimation as well as fault diagnosis of nonlinear systems. Our methodology is based on solving the Bayesian recursive relations through SMC methods. {An on-line parameter estimation scheme is developed based on a modified artificial evolution concept by using the particle filters (PF) approach. Specifically, by using the prediction error to correct the time-varying changes in the system parameters, a novel methodology is proposed for parameter estimation of nonlinear systems based on the PF. Specifically, in the implementation of our proposed scheme, a dual structure for both state and parameter estimation is developed within the PF framework. In other words, the hidden states and variations of the system parameters are estimated through operating two concurrent filters. Convergence and stability properties of our proposed dual estimation strategy are shown to be guaranteed formally under certain conditions.}

The remainder of this paper is organized as follows. In Section \ref{II}, the statement of the nonlinear filtering problem is presented. Our proposed dual state/parameter estimation scheme is developed in Section \ref{IV}, in which state and parameter estimation methods are first developed concurrently and subsequently integrated together for {\textit{simultaneously}} estimating the system states and parameters. The stability and convergence properties of the proposed schemes under certain conditions are also provided in Section \ref{IV}. Our proposed fault diagnosis framework and formulation are also provided in Section \ref{IV}. In Section \ref{V}, extensive simulation results and case studies are provided to demonstrate and justify the merits of our proposed method for fault diagnosis of a gas turbine engine under simultaneous and concurrent component faults. Finally, the paper is concluded in Section \ref{VI}.
\vspace{-2mm}
\section{PROBLEM STATEMENT} \label{II}
\vspace{-0.5mm}
The problem under consideration is to obtain an optimal estimate of states as well as time-varying parameters of a nonlinear system whose dynamics is governed by a discrete-time stochastic model,
\vspace{-2mm}
\begin{eqnarray}\label{states}
x_{t+1}=f_t(x_{t},\theta_{t},\omega_{t}),
\end{eqnarray}
\vspace{-8mm}
\begin{eqnarray}\label{output}
  y_{t}=h_t(x_{t},\theta_{t})+\nu_t,
\end{eqnarray}
where $x_t\in \mathbb{R}^{n_x}$ is the system state, $t\in \mathbb{N}$, $f_{t}:\mathbb{R}^{n_{x}}\times \mathbb{R}^{n_\theta}\times \mathbb{R}^{n_\omega}\longrightarrow \mathbb{R}^{n_{x}}$ is a known nonlinear function, $\theta_{t}\in \mathbb{R}^{n_{\theta}}$ is an {\it{unknown}} and possibly {\it{time-varying}} parameter vector governed by an {\it{unknown dynamics}}. The function $h_{t}:\mathbb{R}^{n_{x}}\times \mathbb{R}^{n_\theta}\longrightarrow \mathbb{R}^{n_{y}}$ is a known nonlinear function representing the map between the states, parameters and the system measurements, and $\omega_{t}$ and $\nu_t$ are uncorrelated stochastic process and measurement noise sequences with covariance matrices $L_t$ and $V_t$, respectively. The following assumption is made regarding the dynamical system \eqref{states} and \eqref{output}.

 {\bf{Assumption A1}}. The vector $\{x_t,\theta_t\}$ ranges over a compact set denoted by $D_{\mathcal{N}}$, for which the functions $f_t(x_t,\theta_t,\omega_t)$ and $h_t({x_t},\theta_t)$ are continuously differentiable with respect to the state $x_t$ as well as the parameter $\theta_t$.

The main objective of the dual state and parameter estimation problem is to approximate the following conditional expectations:
\vspace{-2mm}
\begin{subequations}\label{exp}
\begin{align}
  \mathbb{E}(\phi_1(x_t)|y_{1:t},\theta_{t-1})&=\int \phi_1(x_t)p(x_t|y_{1:t},\theta_{t-1}){\rm{d}}x_t,
  \label{expa}\\
  \mathbb{E}(\phi_2(\theta_t)|y_{1:t},x_t)&=\int \phi_2(\theta_t)p(\theta_t|y_{1:t},x_t){\rm{d}}\theta_t,\label{expb}
 \end{align}
\end{subequations}
where $y_{1:t}=(y_1,y_2,...,y_t)$ denotes the available observations up to time $t$, $\phi_1:\mathbb{R}^{n_x}\rightarrow\mathbb{R}$ and $\phi_2:\mathbb{R}^{n_\theta}\rightarrow\mathbb{R}$ are functions of states and parameters, respectively, that are to be estimated. The conditional probability functions $p(x_t|y_{1:t},\theta_{t-1}){\rm{d}}x_t$ and $p(\theta_t|y_{1:t},x_t){\rm{d}}\theta_t$ are to be approximated by the designed particle filters (PFs) through determining the filtering distributions according to
\vspace{-2mm}
\begin{eqnarray}\label{approx1}
\begin{split}
  \hat{p}_N(x_t|y_{1:t},\theta_{t-1}){\rm{d}}x_t&=\sum_{i=1}^{N}w^{(i)}_{x_t}\delta_{x^{(i)}_t}({\rm{d}}x_t),\\ \hat{p}_N(\theta_t|y_{1:t},x_t){\rm{d}}\theta_t&=\sum_{j=1}^{N}w^{(j)}_{\theta_t}\delta_{\theta^{(j)}_t}({\rm{d}}\theta_t),
  \end{split}
\end{eqnarray}
where the subscript $N$ in $\hat{p}_N(.)$ implies that the state/parameter conditional probability distributions are obtained from $N$ particles. Each state particle $x^{(i)}_t$ has a weight $w^{(i)}_{x_t}$ and each parameter particle $\theta^{(j)}_t$ has a weight $w^{(j)}_{\theta_t}$, where $\delta(.)$ denotes the Dirac-delta function mass that is positioned at $x_t$ or $\theta_t$.

Based on the approximations used in equation \eqref{approx1}, our goal is to address the convergence properties of the subsequently designed estimators to their true optimal estimates and also to develop and demonstrate under what conditions this convergence remains valid.
\vspace{-1mm}
\section{Proposed Dual State/Parameter Estimation and Fault Diagnosis Framework}\label{IV}
In this section, the main theoretical framework for our proposed dual state/parameter filtering as well as the fault diagnosis methodology of the nonlinear system \eqref{states} and \eqref{output} are introduced and developed.
\vspace{-2mm}
\subsection{Dynamic Model in Presence of Time-Varying Parameters} \label{A}
Our first task is to represent the model \eqref{states} and \eqref{output} into another framework for our subsequent theoretical developments. Let $(\Omega,\mathcal{F},P)$ denote the probability space on which the three real vector-valued stochastic processes $X=\{X_t,t=1,2,...\}, \Theta=\{\Theta_t,t=1,2,...\}$, and $Y=\{Y_t,t=1,2,...\}$ are defined. The $n_x$-dimensional process $X$ describes the evolution of the hidden states, the $n_\theta$-dimensional process $\Theta$ describes the evolution of the hidden system parameters that are conditionally independent of the states, and the $n_y$-dimensional process $Y$ denotes the observation process of the system.

The processes $X$ and $\Theta$ are Markov processes with the associated initial state and parameter $X_0$ and $\Theta_0$, respectively. They are drawn from the initial distributions $\pi_{x_0}({\rm{d}}x_0)$ and $\pi_{\theta_0}({\rm{d}}\theta_0)$, respectively. The dynamic evolution of states and parameters are modeled by the Markov transition kernels $K_x({\rm{d}}x_{t}|x_{t-1},\theta_{t-1})$ and $K_\theta({\rm{d}}\theta_{t}|\theta_{t-1},x_{t})$, that also admit densities with respect to the Lebesgue measure {\footnote{ The transition kernel $K({\rm{d}}x_t|x_{t-1})$ admits density with respect to the Lebesgue measure if one can write
$P(X_{t}\in {\rm{d}}x_{t}|X_{t-1}=x_{t-1})=K({\rm{d}}x_{t}|x_{t-1})=K(x_{t}|x_{t-1}){\rm{d}}x_{t}.$}, such that
\vspace{-1mm}
{\small{
\begin{align}\label{kernels}
\begin{split}
P(X_{t}\in A_1|X_{t-1}=x_{t-1},\Theta_{t-1}=\theta_{t-1}) =\int_{A_1}K_x(x_{t}|x_{t-1},\theta_{t-1}){\rm{d}}x_{t},
\end{split}
\end{align}
\vspace{-4mm}
\begin{align}\label{kernels2}
\begin{split}
P(\Theta_{t}\in A_2|\Theta_{t-1}=\theta_{t-1} ,X_{t}=x_{t})=\int_{A_2}K_\theta(\theta_{t}|\theta_{t-1},x_{t}){\rm{d}}\theta_{t},
\end{split}
\end{align} }}for all $A_1\in \mathcal{B}(\mathbb{R}^{n_x})$ and $A_2\in \mathcal{B}(\mathbb{R}^{n_\theta})$, where $\mathcal{B}(\mathbb{R}^{n_x})$ and $\mathcal{B}(\mathbb{R}^{n_\theta})$ denote the Borel $\sigma$-algebra on $\mathbb{R}^{n_x}$ and $\mathbb{R}^{n_\theta}$, respectively. The transition kernel $K_x(x_t|x_{t-1},\theta_{t-1})$ is a probability distribution function (pdf) that follows the pdf of the stochastic process in process \eqref{states}. The probability density function for approximating the parameter kernel transition $K_{\theta}(\theta_t|\theta_{t-1},x_t)$ is to be provided in the subsequent subsections.

Given the states and parameters, the observations $Y_t$ are conditionally independent and have the marginal distribution with a density with respect to the Lebesgue measure as given by,
\vspace{-2mm}
\begin{eqnarray}\label{out_ker}
P(Y_t\in B|X_t=x_t,\Theta_t = \theta_t)&=\int_{B} \rho (y_{t}|x_t,\theta_t){\rm{d}}y_{t},
\end{eqnarray}
where $\rho(y_t|x_t,\theta_t)$ is a probability density function that follows the probability density function of the stochastic process in equation \eqref{output}.

In the dual state/parameter estimation framework, at first the state $x_t$ is estimated (which is denoted by $\hat x_{t|t}$). The estimated value at time $t$ is then used to estimate the parameter $\theta_{t}$ at time $t$ (which is denoted by $\hat \theta_{t|t}$). In the Bayesian framework for parameter estimation, the prior evolution of parameters are not specified, therefore it is necessary to consider a given evolution for the parameters in order to design an estimation filter. In our proposed dual structure for the state estimation filter, first the parameters are assumed to be constant at time $t-1$ at their estimated value $\hat \theta_{t-1|t-1}$, and then for the parameter estimation filter they are evolved to the next time instant by applying an update law that is based on the prediction error (PE) method. The details regarding our proposed methodology are presented in the subsequent subsections in which the filtering of states and parameters are fully described and developed.
\vspace{-2mm}
\subsection{The Dual State/Parameter Estimation Framework}\label{dual_new}
In our proposed dual state/parameter estimation framework, two filters are runing concurrently. At every time step, the first PF-based state filter estimates the states by using the current available estimate of the parameters, $\hat \theta_{t-1|t-1}$, whereas the second PF-based parameter filter estimates the unknown parameters by using the current estimated states, $\hat x_{t|t}$. The developed schematic is shown in Figure \ref{fig:dual_schematic}.
 \begin{figure}
 \vspace{-3mm}
\centering
\includegraphics[scale=0.7]{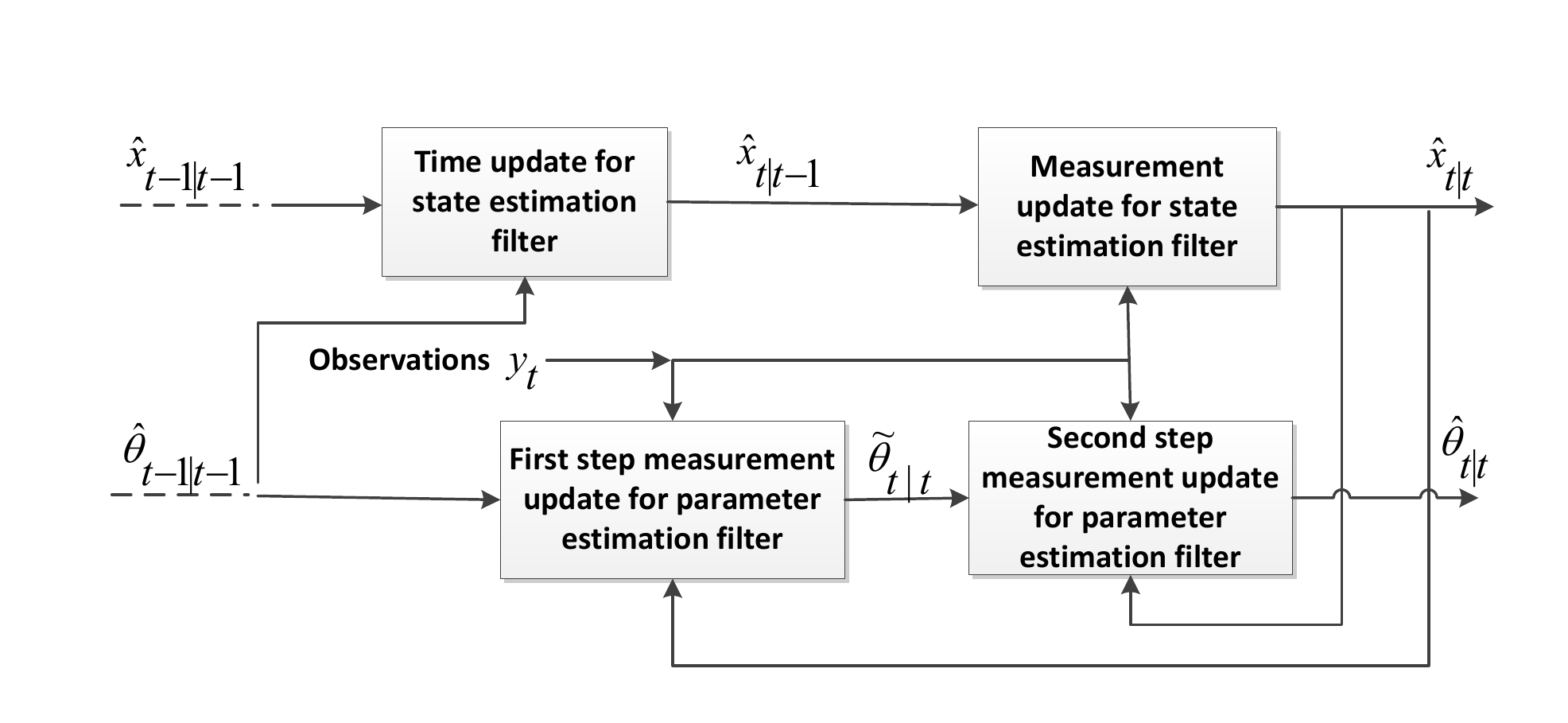}
 \vspace{-10mm}
\caption{The schematic of the dual particle filter.}
\label{fig:dual_schematic}
\end{figure}

In our dual estimation framework, the well-known maximum {\it {a posteriori}} (MAP) solution corresponding to the marginal estimation methods based on the decoupled approach is used for solving the dual estimation problem \cite{Haykin}. In this method, the joint state/parameter marginal density $p(x_t,\theta _t|y_{1:t})$ is expressed as
\begin{eqnarray}\label{dual_main}
p(x_t,\theta _t|y_{1:t})=p(x_t|\theta_{t},y_{1:t})p(\theta_t|y_{1:t}),
\end{eqnarray}
where $p(x_t|\theta_{t},y_{1:t})$ and $p(\theta_t|y_{1:t})$ denote the state and parameter marginal densities, respectively. Assuming that the variations of parameters are slow when compared to the system state time variations, one can use the approximation $\theta_t\approx\theta_{t-1}$, so that the joint marginal density is approximated as
\begin{eqnarray}\label{dual_approx}
p(x_t,\theta _t|y_{1:t})\approx  p(x_t|\theta_{t-1},y_{1:t})p(\theta_{t-1}|y_{1:t}).
\end{eqnarray}
Our ultimate goal is to maximize the two marginal distribution terms in expression \eqref{dual_approx} separately according to the decoupled approach in \cite{Haykin} as follows
{\small{\begin{eqnarray}
\hat x_{t|t} ={\rm{arg max}}_{x_t}p (x_t|\theta _{t-1},y_{1:t}),\;\;\; \hat \theta_{t|t} ={\rm{arg max}}_{\theta_{t-1}}p(\theta_{t-1}|y_{1:t}).
\end{eqnarray}}}

 {In the above decoupled methodology, one attribute is optimized at a time by keeping the other attribute fixed} and then alternating them. Associated with optimization of both marginal distributions, different cost functions can be chosen \cite{Haykin}. For developing a dual extended Kalman filter, corresponding to specific cost functions of the parameter marginal density, various estimation methods have been proposed in the literature \cite{Haykin, Haykin2}. For example, the maximum likelihood (ML) and prediction error approaches are selected for marginal estimations. The main motivation for choosing these two approaches is due to the fact that one considers to maximize only the marginal density $p(\theta_{t-1}|y_{1:t})$ as opposed to the joint density $p(x_t,\theta_{t-1}|y_{1:t})$. However, in order to maximize the parameter marginal density, it is also necessary to generate state estimates that are produced by maximizing the state marginal density $p(x_t|\theta_{t-1},y_{1:t})$.

It should be noted that in marginal estimation methods no explicit cost function is considered for maximization of the state marginal distribution, since the state estimation is only an implicit step in marginal approaches and the joint state/parameter cost is used that may have variety of forms in different filtering algorithms \cite{Haykin}. In our proposed dual particle filtering framework, $p(x_t|\theta_{t-1},y_{1:t})$ is approximated by the state filtering distribution $\hat p_N(x_t|\theta_{t-1},y_{1:t})$ from equation \eqref{approx1}. Next, the prediction error cost function is chosen for maximization of the parameter marginal density, where this cost function is implemented through a PE approach in order to attain a less computational cost \cite{Walid}.

 In the subsequent subsections, specific details regarding the concurrent state and parameter estimation filters design and development are provided.
\vspace{-2mm}
\subsection{The State Estimation Problem}\label{B}
For designing the state and parameter estimation filters, our main objectives are to approximate the integrals in equations \eqref{kernels} and \eqref{kernels2} by invoking the particle filter (PF) scheme as well as to approximate the estimate of the conditional state and parameter distributions. Consider $\pi_{x_{t|t-1}}({\rm{d}}x_t)=\int_{\mathbb{R}^{n_x}}\pi_{x_{t-1|t-1}}({\rm {d}}x_{t-1})K_x({\rm{d}}x_t|x_{t-1},\theta_{t-1})$ to denote the {\it{a priori}} state estimation distribution before the observation at time $t$ becomes available, and $\pi_{\theta_{t-1|t-1}}({\rm{d}}\theta_{t-1})$ to denote the marginal distribution of the parameter at time $t-1$. The {\it{a posteriori}} state distribution after the observation at the instant $t$ becomes available is obtained according to the following rule,
\vspace{-2mm}
\begin{eqnarray}\label{pix}
\pi_{x_{t|t}}({\rm{d}}x_t)\propto \rho(y_t|x_{t},\theta_{t-1})\pi_{x_{t|t-1}}({\rm{d}}x_t)\pi_{\theta_{t-1|t-1}}({\rm{d}}\theta_{t-1}).
\end{eqnarray}
In the above it is assumed that $\hat \theta_{t-1|t-1}$ is known for this filter. Therefore, the last distribution in the right hand side of equation \eqref{pix} is set to one.

The particle filter (PF) procedure for implementation of the state estimation and for determining $\pi_{x_{t|t}}({\rm{d}}x_t)$ consists of \textit{two main steps}, namely {\bf{(a)}} the prediction step (time update step), and {\bf{(b)}} the measurement update step. Consider one states in the $N$ particles at time $t$. The prediction step utilizes the knowledge of the previous distribution of the states as well as the previous parameter estimate, these are denoted by $\{\hat x_{t-1|t-1}^{(i)},\;i=1,...,N\}$ (corresponding to $N$ estimated state particles that follow the distribution $\pi_{x_{t-1|t-1}}({\rm{d}}x_{t-1})$) and $\hat \theta_{t-1|t-1}$, respectively, as well as the process model given by equation \eqref{states}. In other words, the prediction step is explicitly governed by the following equations for $i=1,...,N$, namely
\vspace{-1mm}
\begin{subequations}\label{state}
\begin{align}
\hat x_{t|t-1}^{(i)}&=f_t(\hat x_{t-1|t-1}^{(i)},\hat\theta_{t-1|t-1},{\omega}_t^{(i)}),\label{state_a}\\
{\hat y}_{t|t-1}^{(i)}&=h_t(\hat x_{t|t-1}^{(i)},\hat\theta_{t-1|t-1}),\label{state_b}\\
\Sigma_{\hat x_{t|t-1}}&=(\hat x^{(i)}_{t|t-1}-\frac{1}{N}\sum_{i=1}^N \hat x^{(i)}_{t|t-1})(\hat x^{(i)}_{t|t-1}-\frac{1}{N}\sum_{i=1}^N \hat x^{(i)}_{t|t-1})^{\rm{T}},\label{state_c}
\end{align}
\end{subequations}
where ${\omega}_t^{(i)}$ denotes the process noise related to each particle $\hat x^{(i)}_{t|t-1}$ and is drawn from the noise distribution with the probability distribution function $p_{\omega_t}(.)$, and $\hat x^{(i)}_{t|t-1}$ denotes the independent samples generated from equation \eqref{state_a} for $i=1,...,N$ particles. Moreover, $\hat y^{(i)}_{t|t-1}$ denotes the independent samples of the predicted outputs that are evaluated at $\hat x^{(i)}_{t|t-1}$ samples, and $\Sigma_{\hat x_{t|t-1}}$ denotes the {\it{a priori}} state estimation covariance matrix.

For the \textit{{first step}}, the one-step ahead prediction distribution known as the {\it{a priori}} state estimation distribution is now given by,
\vspace{-4mm}
\begin{equation}
{\tilde \pi}^N_{x_{t|t-1}}({\rm{d}}x_t)\triangleq\frac{1}{N}\sum_{i=1}^{N}\delta_{{\hat x}_{t|t-1}^{(i)}}({\rm{d}}x_t),
\end{equation}

For the \textit{{second step}}, the information on the present observation $y_t$ is used. This results in  approximating $\pi_{x_{t|t}}({\rm{d}}x_t)$, where $\hat \theta_{t-1|t-1}$ is considered to be given from a parameter estimation filter and obtained from the distribution $\pi^N_{\theta_{t-1|t-1}}({\rm{d}}\theta_{t-1})$. Consequently, the particle weights $w^{(i)}_{x_t}$ are updated by the likelihood function (the importance function) according to $w_{x_t}^{(i)}\sim p_{\nu_t}(y_t-{\hat y}_{t|t-1}^{(i)})=\rho(y_t|\hat x^{(i)}_{t|t-1},\hat\theta_{t-1|t-1})$, where $p_{\nu_t}(.)$ denotes the probability distribution function of the additive noise of the output and is evaluated at $y_t-{\hat y}_{t|t-1}^{(i)}$.

In this work, since our ultimate goal is in developing a fault diagnosis algorithm that is practically stable, the structure of regularized particle filters (RPF) is chosen which is quite common in many practical applications \cite{RPF_2002}. The characteristics of the RPFs are related to the fact that they are capable of transforming the discrete-time approximation of the {\it{a posteriori}} state estimation distribution $\pi^N_{x_{t|t}}({\rm{d}}x_t)$ into a continuous-time one. Consequently, the resampling step is modified in such a manner that the new resampled particles are obtained from an absolutely continuous-time distribution with $N$ different locations $\hat x_{t|t}^{(i)}$ from that of $\hat x_{t|t-1}^{(i)}$ \cite{RPF}. Therefore, where the probability for taking the $k$-th particle is $P(\hat x_{t|t}^{(i)}=\hat x^{(k)}_{t|t-1})=\tilde w_{x_t}^{(k)}\triangleq\frac{\rho(y_t|\hat x^{(k)}_{t|t-1},\hat\theta_{t-1|t-1})}{\sum^N_{k=1}\rho(y_t|\hat x^{(k)}_{t|t-1},\hat\theta_{t-1|t-1})}$, and $\tilde w_{x_t}^{(k)}$ for $k=1,...,N$ denotes the normalized particle weights. In other words, the particle selection in the resampling step is performed for particles that have higher probabilities of $\rho_{\nu_t}(y_t-\hat y_{t|t-1}^{(k)})$.

For the resampling step, two main choices can be considered that are known as {\bf{(i)}} Bayesian bootstrap, and {\bf{(ii)}} Sampling importance resampling (SIR) \cite{Orchard_2}. Although both approaches are applicable for this filter, the bootstrap method is chosen in this paper. Therefore, the {\it{a posteriori}} state estimation distribution is approximated by $\tilde \pi^N_{x_{t|t}}({\rm{d}}x_t)$ before one performs the resampling by using the RPF structure \cite{RPF}, and by $\pi^N_{x_{t|t}}({\rm{d}}x_t)$ after one performs the resampling that is provided below,
\vspace{-2mm}
 \begin{eqnarray}\label{posteriori}
\begin{split}
&\tilde \pi^N_{x_{t|t}}({\rm{d}}x_t) \approx \sum_{l=1}^{N_{\rm{reg}}}\sum_{i=1}^{N} \tilde w_{x_t}^{(i)}\frac{|\mathbb{A}_t^{-1}|}{b^{n_x}}\mathcal{K}(\frac{1}{b}\mathbb{A}_t^{-1}(x^{{\rm{reg}}_l}_{t}-\hat x^{(i)}_{t|t-1})),\\
&\tilde w^{(i)}_{x_{t}}\triangleq\frac{\rho(y_t|\hat x^{(i)}_{t|t-1},\hat\theta_{t-1|t-1})}{\sum^N_{i=1}\rho(y_t|\hat x^{(i)}_{t|t-1},\hat\theta_{t-1|t-1})},\\
& \pi^N_{x_{t|t}}({\rm{d}}x_t)=\frac{1}{N}\sum^N_{i=1}\delta_{\hat x^{(i)}_{t|t}}({\rm{d}}x_t)\rightarrow \hat x_{t|t}=\frac{1}{N}\sum^N_{i=1}\hat x^{(i)}_{t|t},
 \end{split}
\end{eqnarray}
where $x^{{\rm{reg}}_l}_{t},\; l=1,..,N_{\rm{reg}}$ denotes the regularized state vector that is evaluated at $N_{\rm{reg}}$ points that are obtained from the absolutely continuous-time distribution of the particles as given by
 \begin{eqnarray}\label{reguralization}
\begin{split}
&X_{t|t-1}=[\hat x_{t|t-1}^{(1)},...,\hat x_{t|t-1}^{(N)}],\\
&x^{{\rm{reg}}_1}_{t}={\rm{min}}(X_{t|t-1})-{\rm{std}}(X_{t|t-1}),\\
&x^{{\rm{reg}}_{N_{\rm{reg}}}}_t={\rm{max}}(X_{t|t-1})+{\rm{std}}(X_{t|t-1}),\\
&{\rm{d}}x^{\rm{reg}}=(x^{{\rm{reg}}_{N_{\rm{reg}}}}_t-x^{{\rm{reg}}_1}_{t})/(N_{\rm{reg}}-1),\\
&x^{{\rm{reg}}_l}_t=x^{{\rm{reg}}_{l-1}}_t+{\rm{d}}x^{\rm{reg}},\; l=2,...,N_{\rm{reg}},
 \end{split}
\end{eqnarray}
where ${\rm{std}}$ denotes the first standard deviation of the particles from their mean. Hence, $\{ \hat x_{t|t}^{(i)}\}^N_{i=1}$ is obtained from the continuous-time distribution through the regularization kernel $\mathcal{K}$ that is considered to be a symmetric density function on $\mathbb{R}^{n_{x}}$ \cite{RPF}. The matrix $\mathbb{A}_t$ in equation \eqref{posteriori} is chosen to yield a unit covariance value in the new $\hat x^{(i)}_{t|t}$ population and $\mathbb{A}_t\mathbb{A}_t^{\rm{T}}=\Sigma_{\hat x_{t|t-1}}$. The constant $b$ denotes the optimal bandwidth of the kernel, and $\hat x_{t|t}$ denotes the {\it{a posteriori}} state estimation at time $t$.

We are now in a position to introduce our overall particle filter (PF) scheme for implementing the state estimation filter. Our goal for proposing this algorithm is to ensure that an approximation to $\mathbb{E}(\phi(x_t)|y_{1:t},\theta_{t-1})$ by $\phi(x_t)=x_t$ takes $\hat x_{t|t}\sim \pi^{N}_{x_{t|t}}({\rm{d}}x_t)=\frac{1}{N}\sum_{i=1}^{N}\delta_{\hat x^{(i)}_{t|t}}({\rm{d}}x_t)$, where $\pi^{N}_{x_{t|t}}({\rm{d}}x_t)$ denotes the {\it{a posteriori}} distribution of $\{\hat x^{(i)}_{t|t}\}^{N}_{i=1}$ (after the resampling from $\{\hat x^{(i)}_{t|t-1}\}^{N}_{i=1}$), that is given by $\hat x_{t|t}=\frac{1}{N}\sum^N_{i=1}\hat x^{(i)}_{t|t}$. The estimated output from the state estimation filter is also given by $\hat y_t= h_t(\hat x_{t|t},\hat\theta_{t-1|t-1})$.
\vspace{-1mm}
\subsection*{\underline{The State Estimation Particle Filter Scheme}}
\begin{enumerate}
\item Initialize the PF scheme with $N$ particles, $\{x^{(i)}_0\}_{i=1}^{N}\thicksim \pi_{x_0}({\rm{d}}x_0)$ and the parameters $\theta_0$ (the mean of the parameter initial distribution $\pi_{\theta_0}({\rm{d}}\theta_0)$).
\item Draw $\omega_t^{(i)}\sim p_{\omega_t}(.)$, where $p_{\omega_t}(.)$ denotes a given distribution for the process noise in the filter, and then predict the state particles $\hat x_{t|t-1}^{(i)}$ according to equation \eqref{state_a}.
\item Compute $\hat y^{(i)}_{t|t-1}$ from equation \eqref{state_b} to obtain the importance weights $\{ w^{(i)}_{x_t} \}^{N}_{i=1}$ as $w_{x_t}^{(i)}=\rho(y_t|\hat x^{(i)}_{t|t-1},\hat\theta_{t-1|t-1}),\; i=1,...,N,$ and normalize them to $\tilde{w}^{(i)}_{x_t}=\frac{w^{(i)}_{x_t}}{\sum ^N_{i=1}w^{(i)}_{x_t}}$.
\item Resampling: Draw $N$ new particles with the replacement for each $i=1,...,N$, according to $P(\hat x^{(i)}_{t|t}=\hat{x}^{(k)}_{t|t-1})=\tilde{w}^{(k)}_{x_t},\; k=1,...,N$, from the regularized kernel $\mathcal{K}$ where $\hat x^{(i)}_{t|t}\sim \tilde \pi_{x_{t|t}}^N({\rm{d}}x_t)$ as given by equation \eqref{posteriori}.
\item Calculate $\hat x_{t|t}$ from the conditional distribution that is given by equation \eqref{posteriori},\\ $\pi^N_{x_{t|t}}({\rm{d}}x_t)=\frac{1}{N}\sum^N_{i=1}\delta_{\hat x^{(i)}_{t|t}}({\rm{d}}x_t)$ with equally weighted $\hat x^{(i)}_{t|t}$ as $\hat x_{t|t}=\frac{1}{N}\sum^N_{i=1}\hat x^{(i)}_{t|t}$.
\item Update the parameters from the parameter estimation filter (to be specified in the next subsection).
\item Set $t:=t+1$ and go to Step $2$.
\end{enumerate}

Following the implementation of the above state estimation filter, the parameter estimation filter that is utilized for adjusting the parameters is now described in detail in the next subsection.
\vspace{-2mm}
\subsection{The Parameter Estimation Problem}\label{D}
One of the main contributions of this paper is to develop a novel PF-based parameter estimation filter within our proposed dual state/parameter estimation framework by utilizing the prediction error (PE) concept. For this methodology it is assumed that the {\it{a priori}} distribution of the time-varying developed is not known. Moreover, the estimated states that are generated by the state estimation filter provided in the previous subsection will be used. Therefore, it is imperative that one considers a dynamical model associated with the parameters evolution in order to estimate the density function $\pi _{\theta_{t|t}}({\rm{d}}\theta_t)$.

The most common dynamical model that is considered for the parameter propagation (in case of the system with constant parameters) is the conventional artificial evolution law. In this representation small random disturbances are added to the state particles (parameters) between each consecutive time step \cite{Parameter}. However, in our work, the conventional update law for the parameters is modified to include the output prediction error as a multiplicative term to allow one to deal with time variations in the parameters that can affect the system output.

In order to derive the parameter update law, an algorithm based on the prediction error (PE) method is proposed by minimizing the expectation of a quadratic performance index $\bar J(\theta_{t-1})$ with respect to $ \theta_{t-1}$. This is due to the fact that our parameter estimation algorithm for obtaining the distribution of the {\it{a posteriori}} parameter estimate is based on the kernel smoothing that uses the shrinkage of the particle locations. This method attempts to force the particles towards their mean from the previous time step, i.e. the estimated value of $\theta_{t-1}$, and is denoted by $\hat \theta_{t-1|t-1}$ (before adding noise to the particles). This is also used in the state estimation filter for approximating $\hat x_{t|t}$. Therefore, our goal is to investigate the convergence properties of $\hat \theta_{t-1|t-1}$ whose boundedness ensures the boundedness of $\hat \theta_{t|t}$. Towards this end, the performance index is now selected as $\mathbb{E}(\bar J(\theta_{t-1})|y_{1:t-1}, x_{t})=\int \bar J(\theta_{t-1})p(\theta_{t-1}|y_{1:t-1},x_{t}){\rm{d}}\theta_{t-1},$ where the integral is approximated in the PF by $\mathbb{E}(\bar J(\theta_{t-1})|y_{1:t-1}, x_{t})\approx\frac{1}{N}\sum^N_{j=1}\bar J(\hat \theta^{(j)}_{t-1|t-1})$.

The term $\bar J(\hat \theta^{(j)}_{t-1|t-1})$ now represents a quadratic function of the output prediction error related to each particle $j,\;j=1,...,N$. The prediction error is now defined according to $\epsilon(t,\hat\theta^{(j)}_{t-1|t-1})\triangleq\epsilon_t^{(j)}=y_t-h_t(\hat x_{t|t},\hat \theta^{(j)}_{t-1|t-1})$, where $\hat \theta^{(j)} _{t-1|t-1}$ denotes the particle related to the estimated value of the parameter whose true value is denoted by $\theta^{\star}_{t-1}$ (this is clearly assumed to be unknown). Therefore, we define $\bar J(\hat\theta^{(j)}_{t-1|t-1})=\frac{1}{\kappa}\sum_{\tau=t-\kappa}^{\tau=t}\mathbb{E}(Q(\epsilon(\tau,{\hat\theta^{(j)}_{t-1|t-1}})))$, in which the expectation is taken over the observation sequence of $\kappa$ samples. Let us now select the quadratic criterion $Q(\epsilon(t,\hat \theta^{(j)}_{t-1|t-1}))$ as
\vspace{-3mm}
  \begin{eqnarray}\label{l}
  Q(\epsilon(t,\hat \theta^{(j)}_{t-1|t-1}))=\frac{1}{2}\epsilon(t,\hat \theta^{(j)}_{t-1|t-1})\epsilon ^{\rm{T}}(t,\hat \theta^{(j)}_{t-1|t-1}).
 \end{eqnarray}

The following modified artificial evolution law is now proposed for the parameter update in the particle filters for generating $j=1,...,N$ parameter particles that correspondingly determine the distribution from which the {\it{a priori}} parameter estimate $\hat \theta_{t|t-1}^{(j)}$ is considered to be the same as $\hat \theta_{t-1|t-1}^{(j)}$, and the {\it{a posteriori}} parameter estimate is obtained in two steps that are denoted by $\tilde \theta^{(j)}_{t|t}$ and $\hat \theta_{t|t}^{(j)}$, respectively.

In the {\textit{first step}}  (the second step is described in the next page)  for $ \tilde \theta^{(j)}_{t|t} $   one gets
\vspace{-2mm}
 \begin{subequations}
 \begin{align}
&m^{(j)}_t = \hat\theta^{(j)}_{t-1|t-1}+\gamma_t R^{(j)}_t\psi^{(j)} _t\epsilon(t,\hat \theta^{(j)}_{t-1|t-1}),\label{descent3}\\
 &\tilde \theta^{(j)}_{t|t}= Am^{(j)}_t+(I-A)\bar m_{t-1}+\zeta^{(j)}_t,\;  \bar m_{t-1}=\frac{1}{N}\sum^N_{j=1}\hat \theta_{t-1|t-1}^{(j)}, \label{descent2}
  \end{align}
  \end{subequations}
where $\psi _t=\frac{{\partial}\hat y_{t}}{{\partial}\hat \theta_{t-1|t-1}}=\frac{{\partial}h_{t}(\hat x_{t|t},\hat \theta_{t-1|t-1})}{{\partial}{\hat \theta_{t-1|t-1}}}$, which when evaluated at $\hat \theta_{t-1|t-1}^{(j)}$ is denoted by $\psi^{(j)} _t$, $\gamma_t$ denotes the step size design parameter, $\zeta^{(j)}_{t}\sim \mathcal{N}(0,(I-A^2)V_{\hat \theta_{t-1|t-1}})$ denotes the zero-mean normal increment particles to the parameter update law at each time step with the covariance matrix $(I-A^2)V_{\hat \theta_{t-1|t-1}}$ through the use of the kernel smoothing concept, $A$ denotes the shrinkage matrix, and $V_{\hat \theta _{t-1|t-1}}$ denotes the covariance of the parameter estimates in the previous time step $t-1$. The kernel shrinkage algorithm attempts to force the parameter particles distribution  towards its mean in the previous time instant that was denoted by $\bar m_{t-1}$, by applying the shrinkage coefficient matrix $A$ to the obtained $m_t^{(j)}$.

The processes $\hat{\theta}^{(j)}_{t-1|t-1}$ and $\zeta^{(j)}_t$ are conditionally independent given observations up to time $t$. Moreover, $R^{(j)}_t=\sqrt{{\rm{trace}}(\mathcal E^{(j)}_t \mathcal E^{(j)^{\rm{T}}}_t)}$ where $\mathcal E^{(j)}_t = \epsilon_t(\hat \theta_{t-1|t-1}^{(j)})-\frac{1}{n_y}\sum_{l=1}^{l=n_y}\epsilon^{(l)}_t(\hat \theta_{t-1|t-1}^{(j)})$ and $\epsilon^{(l)}_t(\hat \theta_{t-1|t-1}^{(j)})$ denotes the $l$-th element of the vector $\epsilon_t(\hat \theta_{t-1|t-1}^{(j)})$. The term $R_t^{(j)}$ denotes a time-varying coefficient that determines the updating direction and is a positive scalar to ensure that the criterion \eqref{l} can be minimized by changing $m^{(j)}_{t}$ in the steepest descent direction. Therefore, the first step estimate of the {\it{a posteriori}} parameter estimation particle is denoted by $\tilde \theta_{t|t}^{(j)}$. The convergence of the update law \eqref{descent3}-\eqref{descent2} will be shown in the Subsection \ref{F}.

The parameter update law according to \eqref{descent3}-\eqref{descent2} contains a term in addition to the independent normal increment $\zeta_t^{(j)}$. The estimated parameter from this update law is invoked in the PF-based parameter estimation filter to represent the distribution from which the parameter particle population for the next time step is chosen. Therefore, the above proposed prediction error based modified artificial evolution law enables the PF-based estimation algorithm to handle and cope with the time-varying parameter scenarios. The time-varying term $\gamma_t R^{(j)}_t$ acts as an {\textit{adaptive step size}} in equations \eqref{descent3}-\eqref{descent2}, and therefore our algorithm can also be considered as an adaptive step size scheme.

In order to ensure that the obtained $\tilde{\theta}^{(j)}_{t|t}$ from the modified artificial evolution law given by equations \eqref{descent3}-\eqref{descent2} remains in $D_\mathcal{N}$ (refer to Assumption A1), the following projection algorithm is utilized that forces $\tilde{\theta}^{(j)}_{t|t}$ to remain inside $ D_\mathcal{N}$ according to the following procedure \cite{Walid},
  \begin{enumerate}
 \item{Choose a factor $0\le\mu\le1$},
 \item {Compute $\breve {\theta}_{t|t}^{(j)}:=\gamma_tR_t^{(j)}\psi_t^{(j)}\epsilon(t,\hat \theta_{t-1|t-1}^{(j)})$},
 \item{Construct $m^{(j)}_{t} :=\hat{\theta}_{t-1|t-1}^{(j)}+\breve {\theta}_{t|t}^{(j)}$},
 \item{If $m^{(j)}_{t}\in D_\mathcal{N}$ go to Step $6$, else go to Step $5$},
 \item{Set $\breve {\theta}_{t|t}^{(j)}=\mu\breve {\theta}_{t|t}^{(j)}$, and go to Step 3},
 \item{Stop.}
 \end{enumerate}

It should be noted that the main reason for considering the above mapping is related the fact that the actual dynamics of the parameters are not known, therefore such mapping ensures that the assumed dynamics for the parameters based on modified artificial evolution model does not cause instability of the entire system.

Consequently, the {\it{a priori}} distribution of the parameter $\theta_t$ is assumed to have the same distribution as in the previous time step. On the other hand, as the present observation $y_t$ becomes available in the measurement update step, the {\it{a posteriori}} distribution of the parameter is obtained through two steps that denoted by ${\tilde\pi}^N_{\tilde \theta_{t|t}}({\rm{d}}\theta_t)$ and ${\tilde\pi}^N_{\theta_{t|t}}({\rm{d}}\theta_t)$, respectively. In what follows, more details related to these distributions are presented.

Consider equations \eqref{descent3}-\eqref{descent2}. The  {\it{a posteriori}} distribution of the parameters calculated from the distribution of the parameter particles $\tilde \theta_{t|t}^{(j)}$ is given by,
\begin{eqnarray}
&{\tilde\pi}^N_{\tilde \theta_{t|t}}({\rm{d}}\theta_t)\triangleq\frac{1}{N}\sum_{j=1}^{N}\delta_{{\tilde \theta}_{t|t}^{(j)}}({\rm{d}}\theta_t),
\end{eqnarray}
and the measurement equation is expressed as,
\begin{eqnarray}
 \bar {y}_{t|t}^{(j)}=h_t(\hat x_{t|t}, {\tilde\theta}_{t|t}^{(j)}),\label{parfilterc}
\end{eqnarray}
where $\bar{y} _{t|t}^{(j)}$ denotes the evaluated output that is obtained by the parameter estimation filter that is different from the one that is obtained by the state estimation filter, as provided in the Subsection \ref{B}.

Now, {in the \textit{second step} for estimating the {\it{a posteriori}} parameter estimate distribution, consider the present observation $y_t$,} so that the particle weights $w^{(j)}_{\theta_t}$ are updated by the likelihood function according to $w_{\theta_t}^{(j)}\sim p_{\nu_t}(y_t-{\bar y}_{t|t}^{(j)})=\rho(y_t|\hat x_{t|t},\tilde\theta^{(j)}_{t|t})$. This can now be expressed by using the normalized weights $\tilde{w}^{(j)}_{\theta_t}$ as
${\tilde\pi}^{N}_{\theta_{t|t}}({\rm{d}}\theta_t)=\sum_{j=1}^N \tilde{w}^{(j)}_{\theta_t} \delta_{{\tilde\theta}^{(j)}_{t|t}}({\rm{d}}\theta_t)$, where $ \tilde{w}^{(j)}_{\theta_t} \triangleq \frac {\rho(y_t|\hat x_{t|t},{\tilde\theta}^{(j)}_{t|t})}{\sum_{j=1}^{N}\rho(y_t|\hat x_{t|t},{\tilde\theta} ^{(j)}_{t|t})}$. Following the resampling/selection step, an equally weighted particle distribution $\pi^N_{\theta_{t|t}}({\rm{d}}\theta_t)$ is obtained as $\pi^N_{\theta_{t|t}}({\rm{d}}\theta_t)=\frac{1}{N}\sum_{j=1}^N \delta_{\hat \theta^{(j)}_{t|t}}({\rm{d}}\theta_t)$ for approximating $\pi_{\theta_{t|t}}({\rm{d}}\theta_t)$, and the resampled (selected) particles that are denoted by $\hat\theta^{(j)}_{t|t}$ follow the distribution $\tilde{\pi}^N_{\theta_{t|t}}({\rm{d}}\theta_t)$. Therefore, the {\it{a posteriori}} parameter estimation distribution is approximated by a weighted sum of the Dirac-delta masses as $\tilde \pi^N_{\theta_{t|t}}({\rm{d}}\theta_t)$ before one performs the resampling and with an equally weighted particle distribution approximation as $\pi^N_{\theta_{t|t}}({\rm{d}}\theta_t)$ according to
\vspace{-3mm}
 \begin{eqnarray}\label{posteriori_2}
\begin{split}
&\tilde \pi^N_{\theta_{t|t}}({\rm{d}}\theta_t) \approx \sum_{j=1}^{N} \tilde w_{\theta_t}^{(j)}\delta_{\tilde \theta ^{(j)}_{t|t}}({\rm{d}}\theta _t),\\
&\tilde w^{(j)}_{\theta_{t}}\triangleq\frac{\rho(y_t|\hat x_{t|t},\tilde \theta^{(j)}_{t|t})}{\sum^N_{j=1}\rho(y_t|\hat x_{t|t},\tilde \theta^{(j)}_{t|t})},\\
& \pi^N_{\theta_{t|t}}({\rm{d}}\theta_t)=\frac{1}{N}\sum^N_{j=1}\delta_{\hat \theta^{(j)}_{t|t}}({\rm{d}}\theta_t)\rightarrow \hat \theta_{t|t}=\frac{1}{N}\sum^N_{j=1}\hat \theta^{(j)}_{t|t},
 \end{split}
\end{eqnarray}
where $\tilde w^{(j)}_{\theta_{t}}$ denotes the normalized parameter particle weight, $\{ \hat \theta_{t|t}^{(j)}\}^N_{j=1}$ is obtained from the resampling/selection step of the scheme by duplicating the particles $\tilde \theta^{(j)}_{t|t}$ having large weights and discarding the ones with small values to emphasize the zones with higher {\it{a posteriori}} probabilities according to $P(\hat \theta ^{(j)}_{t|t}=\tilde \theta ^{(k)}_{t|t})=\tilde w_{\theta _t}^{(k)},\; k=1,...,N$. In our proposed filter the residual resampling method is used to ensure that the variance reduction among the resampled particles is guaranteed \cite{residual}.

 Therefore, an approximation to $\mathbb{E}(\phi(\theta_t)|y_{1:t},x_t)$ by $\phi(\theta_t)=\theta_t$ takes on the form $\hat \theta_{t|t}\sim \pi^N_{\theta_{t|t}}({\rm{d}}\theta_t)=\frac{1}{N}\sum^N_{j=1}\delta_{\hat \theta_{t|t}^{(j)}}({\rm{d}}\theta_t)$, where $\pi^N_{\theta_{t|t}}({\rm{d}}\theta_t)$ denotes the {\it{a posteriori}} distribution of the parameter estimate (after performing the resampling from $\tilde \theta^{(j)}_{t|t}$). The resulting estimated output of this filter is obtained by $\hat y_t=h_t(\hat x_{t|t},\hat \theta_{t|t})$. The explicit details for implementation of the parameter estimation filter are now provided below.

\vspace{-1.5mm}
\subsection*{\underline{The Parameter Estimation Filter}}
The particle filter for implementation of the parameter estimation is described as follows:
\begin{enumerate}
\item Initialize the $N$ particles for the parameters as $\{\theta^j_0\}_{j=1}^{N}\thicksim \pi_{\theta_0}({\rm{d}}\theta_0)$, and use the initial values of the states as $x_0$ that represents the mean of the states initial distribution $\pi_{x_0}({\rm{d}}x_0)$.
\item Draw $\zeta_t^{(j)}\sim \mathcal{N}(0,(I-A^2)V_{\hat \theta_{t-1|t-1}})$.
\item Predict $\tilde \theta ^{(j)}_{t|t},\; j=1,...,N$ from equations \eqref{descent3}-\eqref{descent2} with the projection algorithm.
\item Compute the importance weights $\{w^{(j)}_{\theta_t}\}^{N}_{j=1}, w^{(j)}_{\theta_t}=\rho(y_t|\hat x_{t|t},\tilde\theta^{(j)}_{t|t}), j=1,...,N$, and normalize them to $\tilde{w}^{(j)}_{\theta_t}=\frac{w^{(j)}_{\theta_t}}{\sum^N_{j=1}w^{(j)}_{\theta_t}}$.
\item Resampling: Draw $N$ new particles with replacement for each $j=1,...,N$, $P(\hat \theta^{(j)}_{t|t}=\tilde{\theta}^{(k)}_{t|t})=\tilde{w}^{(k)}_{\theta_t},\ k=1,...,N$, where $\tilde \theta^{(j)}_{t|t} \sim \tilde \pi^N_{\theta_{t|t}}({\rm{d}}\theta_t)=\sum^N_{j=1}\tilde w^{(j)}_{\theta_t}\delta_{\tilde \theta^{(j)}_{t|t}}({\rm{d}}\theta_t)$.
\item Construct $\hat \theta_{t|t}$ from the conditional distribution $\pi^N_{\theta_{t|t}}({\rm{d}}\theta_t)=\frac{1}{N}\sum^N_{j=1}\delta_{\hat \theta^{(j)}_{t|t}}({\rm{d}}\theta_t)$ with equally weighted $\hat \theta^{(j)}_{t|t}$ as $\hat \theta_{t|t}=\frac{1}{N}\sum^N_{j=1}\hat \theta^{(j)}_{t|t}$.
\item Set $t=t+1$ and go to Step $2$ of the state estimation filter as provided in the Subsection \ref{B}.
\end{enumerate}

As stated earlier, the kernel from which the parameter particles i.e. $\tilde \theta_{t|t}^{(j)}$ for the next time step is chosen is a Gaussian kernel and its mean is obtained from $m_{t}^{(j)}$ and its variance is obtained based on the kernel smoothing consideration that is provided in the next Subsection \ref{kernel_smoothing}. In the subsections below, the required conditions for boundedness of the parameter transition kernel $K_\theta(.)$ are also investigated and developed.
\vspace{-1mm}
\subsection{Kernel Smoothing (KS) of the Parameters}\label{kernel_smoothing}
In this subsection, the kernel smoothing (KS) approach \cite{west} is utilized to ensure that the variance of the normal distribution which is obtained according to the modified artificial evolution law for the parameter estimates remains bounded.

Consider the modified artificial evolution law \eqref{descent3}-\eqref{descent2} in which $\zeta_t^{(j)}$ is a normal zero-mean uncorrelated random increment to the parameter that is estimated at time $t-1$. If $A=I$, i.e. when there is no kernel shrinkage, as $t\rightarrow \infty$, the variance of the added evolution increases and can therefore yield $\tilde\theta^{(j)}_{t|t}$ in \eqref{descent2} completely unreliable. This phenomenon is known as the loss of information that can also occur between two consequent sampling times \cite{west}. On the other hand, since $\theta_{t}$ is time-varying, generally there will not exist an optimal value for the variance of the evolution noise $\zeta^{(j)}_t$ that remains suitable for all times.

Consequently, the idea of the kernel shrinkage has been proposed in \cite{west} and later updated in \cite{west_2011}. In the kernel shrinkage approach \cite{Parameter}, for the next time step one takes the mean of the estimated parameter distribution in the particle filter according to the following normal distribution
\begin{eqnarray}\label{theta_kernel}
K_{\theta}({\rm{d}} \theta_t|\theta^{(j)}_{t-1},x_t) &\sim &\mathcal{N}(A m_t^{(j)}+(I-A)\bar m_{t-1},({I-A^2}) \nonumber \\
& & V_{\hat\theta_{t-1|t-1}}),
\end{eqnarray}
where $m_t^{(j)}$ for $j=1,...,N,$ is obtained from \eqref{descent3}. By utilizing this kernel shrinkage rule, the resulting normal distribution retains the mean $\bar m_{t-1}$ and has the appropriate variance for avoiding over-dispersion relative to the {\it{a posteriori}} sample. The kernel shrinkage forces the parameter samples towards their mean before the noise $\zeta_t^{(j)}$ is added. In our proposed approach the changes due to the parameter variations are considered in the mean of the parameter estimate distribution through the modified artificial evolution rule. Consequently, the mean of the distribution, i.e. $\bar m_{t-1}$, itself is time-varying and the kernel shrinkage ensures a smooth transition in the estimated parameters even when they are subjected to changes. To eliminate the information loss effect, by taking the variance from both sides of equation \eqref{descent2} results in $V_{\hat\theta_{t|t-1}}=A^2 V_{\hat\theta_{t-1|t-1}}+(I-A^2)V_{\hat\theta_{t-1|t-1}}=V_{\hat\theta_{t-1|t-1}}$. This ensures that the variance of the added random evolution would not cause over-dispersion in the parameter estimation algorithm for all time.

The following proposition specifies an upper bound on the shrinkage factor. This upper bound is calculated in the worst case, that is when the parameter is considered to be constant but the modified evolution law \eqref{descent3}-\eqref{descent2} is used in the parameter estimation filter for estimating it. Utilization of this upper bound in the kernel shrinkage algorithm ensures the boundedness of the variance of the estimated parameters distribution that is obtained according to the PE-based artificial evolution update law and the kernel smoothing augmented with the shrinkage factor.\\
{\it{\bf{Proposition 1:}} Upper bound on the kernel shrinkage factor}: Given the parameter update law \eqref{descent3}-\eqref{descent2}, the estimated parameters conditional normal distribution based on the kernel smoothing as given by equation \eqref{theta_kernel}, results in an upper bound for $A$ that is obtained as $A\le I(1 - \sqrt{\frac{\sigma_{\rm {min}}(P^{2}_{\rm{max}}\Psi V_y\Psi^{\rm{T}}V_{\hat\theta}^{-1})}{{\sigma_{\rm {max}}(P^{2}_{\rm{max}}\Psi V_y\Psi^{\rm{T}}V_{\hat\theta}^{-1})}}})$, where $\Psi$ denotes the $\psi_t^{(j)}$ in equation \eqref{descent3} but considered as a constant parameter between the time steps $t$ and $t-1$. Moreover, $\sigma_{\rm {min}}$ and $\sigma_{\rm {max}}$ denote the minimum and the maximum eigenvalues of a matrix, respectively, $W$ denotes the upper-bound on the variance of the added noise $W_t$, $V_y$ denotes the upper-bound on the variance of the measurement noise $R_t$, $V_{\hat \theta}$ denotes the variance of the parameters when they are constant that can be assumed to be the same as the initial covariance of the parameters, and $P_{\rm{max}}=\gamma_0 \sqrt{\rm{trace}(\mathcal{E_{\rm{max}}}\mathcal{E_{\rm{max}}}^{\rm{T}}})$, where $\gamma_0$ denotes the initial value of the step size, and $\mathcal{E_{\rm{max}}}\mathcal{E_{\rm{max}}}^{\rm{T}}$ is a design parameter that denotes the maximum acceptable variance among the prediction error vector elements. \\
{\bf{Proof:}} The proof  is provided in the appendix.$\blacksquare$

The convergence of the estimated parameter particles $\hat \theta^{(j)}_{t-1|t-1},\;j=1,...,N$ to the local minimum of $\mathbb{E}(\bar J(\hat \theta^{(j)}_{t-1|t-1})| y_{1:t-1},\hat x_{t})$ is now investigated in the following subsection. The developed convergence proof does not ensure the convergence of the PE-based parameter estimation method to the true parameter value, but only to a set of zeros of the gradient of the chosen performance index. The conditions under which the convergence of the estimated parameters to their optimal values can be guaranteed as $N\rightarrow \infty$ is stated in Remark 1 (refer to the next page).
\vspace{-2mm}
\subsection{Convergence of the PE-based Parameter Update Law}\label{F}
In this subsection, it will be shown that the update law \eqref{descent3}-\eqref{descent2} can guarantee the convergence of the parameter estimate particles $\hat \theta^{(j)}_{t-1|t-1},\;j=1,...,N$ (after the resampling step), to a local minimum of $\mathbb{E}(\bar J(\hat\theta^{(j)}_{t-1|t-1})|y_{1:t-1}, \hat x_{t})$, that is located in a compact set of $\{x_t,\theta_t\}$, denoted by $D_{\mathcal N}$ as per Assumption A1.

In order to investigate the convergence of our proposed PE-based modified artificial evolution law for updating the parameter particles distribution and to achieve a local minimization of $\mathbb{E}(\bar J(\theta_{t-1})|y_{1:t-1},x_{t})$, consider equation \eqref{descent3}, where $\gamma_t$ denotes a time-varying step size such that $\lim_{t\rightarrow \infty} \gamma_t =\mu_0>0$, where $\mu_0$ is a small positive constant. The introduction of the step size $\gamma_t$ is necessary to transform the discrete-time model \eqref{descent3}-\eqref{descent2} into a continuous-time representation as shown subsequently.

First, we need to state the following two assumptions A2-A3 according to \cite{Walid}, to guarantee the convergence of our proposed algorithm as presented in our main result below in Theorem 1. Specifically, we have:

{\bf{Assumption A2}}. The function $Q(\epsilon(t,\hat\theta_{t-1|t-1}))$ is sufficiently smooth and twice continuously differentiable w.r.t. $\epsilon$, and $|Q_{\epsilon\epsilon}(\epsilon(t,\hat \theta_{t-1|t-1}))|\leq C$ for $\hat \theta_{t-1|t-1}\in D_{\mathcal{N}}$, where $Q_{\epsilon\epsilon}(\epsilon(t,\hat \theta_{t-1|t-1}))$ denotes the second derivative of $Q(\epsilon(t,\hat \theta_{t-1|t-1}))$ w.r.t. $\epsilon$.

{\bf{Assumption A3}}. The observation sequence $y_t$ (generated from equation \eqref{output}), is such that $\bar{\mathbb E} (Q(\epsilon(t,\hat \theta_{t-1|t-1})))=\bar J(\hat \theta_{t-1|t-1})$ and $\bar{\mathbb E}[\frac{\rm{d}}{{\rm{d}}\hat \theta_{t-1|t-1}}Q(\epsilon(t,\hat \theta_{t-1|t-1}))]=-g(\hat \theta_{t-1|t-1})$ exist for all $\hat \theta_{t-1|t-1} \in D_\mathcal{N}$, where ${\bar{\mathbb{E}}(Q(\epsilon(t,\hat \theta_{t-1|t-1})))=
\frac{1}{\kappa}\sum_{\tau=t-\kappa}^t\mathbb{E}Q(\epsilon(\tau,\hat \theta_{t-1|t-1}))}$.

It must be noted that the kernel shrinkage method, as stated earlier, attempts to retain the mean of the parameter estimation particles at time $t$ near the estimated parameter in the previous time step $t-1$, i.e. $\hat \theta_{t-1|t-1}$. Therefore, in the following theorem the convergence properties of $\hat \theta_{t-1|t-1}$ is addressed. The main result of this section is stated below.
\vspace{-1mm}
\begin{theorem}
Consider the parameter estimation algorithm as specified by the equations \eqref{descent3}-\eqref{parfilterc}. Also consider the {\it{a posteriori}} parameter estimate as governed by equation \eqref{posteriori_2}. Let Assumptions A1 to A3 hold. It now follows that the particles $\hat \theta^{(j)}_{t-1|t-1},\;j=1,...,N$, and consequently the distribution of the estimated parameter particles approximated by the particle filter $\pi_{\theta_{t-1|t-1}}^{N}({\rm{d}}\theta_{t-1})$, w.p.1 converge either to the set $D_{\mathcal{C}}=\{\hat \theta^{(j)}_{t-1|t-1}|\hat \theta^{(j)}_{t-1|t-1} \in D_{\mathcal{N}},\;  \frac{\rm{d}}{{\rm{d}}\hat\theta^{(j)}_{t-1|t-1}}\bar{J}(\hat\theta^{(j)}_{t-1|t-1})\\ =0,\;j=1,...,N\}$ or to the boundary of $D_{\mathcal{N}}$ as $t\rightarrow \infty$.
\end{theorem}
\vspace{-2mm}
{\bf{Proof:}} The proof  is provided in the appendix.$\blacksquare$

The main reason that our proposed dual state/parameter estimation method for its implementation does not necessarily need more particles than the one that we needed for only the state estimation scheme, is illustrated by the result that one can extract and obtain from Theorem 1. According to this theorem, PE-based modified artificial evolution law enables each single particle to tend to $D_{\mathcal{C}}$. Therefore, even increasing the number of particles would not affect the convergence properties of the filter but it can certainly result in a more accurate state/parameter estimates.

{It was indicated earlier that the above theorem can only guarantee boundedness of the estimated parameter distribution from the particle filters and not its convergence to the optimal distribution. However, from practical considerations it is not readily feasible to obtain an exact dynamical behavior that fully specifies variations of the system health parameters. This is due to the fact that these parameters are affected by fault occurrences and/or damage during the operation of the system. Consequently, the estimated parameters optimal values cannot be guaranteed unless Assumptions A1, A2, and A3 are satisfied. However, based on the results of Theorem 1 and Proposition 1 one can ensure that the probability density function and its related kernel $K_{\theta}({\rm{d}}\theta_t|\theta_{t-1},x_{t-1})$ (in the particle filter) do remain bounded. Moreover, the convergence of the dual state and parameter estimation algorithm can be investigated according to the result of Theorem 3.1 in  \cite{shon} as explained in the following remark}.

\vspace{-2mm}
\begin{remark}
Using the extended setting  introduced in \cite{shon}, and assuming that $\rho(y_t|x_t,\hat \theta_t)<\infty,$ and $ K_x(x_t|x_{t-1},\hat \theta_{t-1})<\infty$, the boundedness of the parameter estimation transition kernel $K_{\theta}(\hat \theta_t|\hat \theta_{t-1},\hat x_t)$ is  ensured from the Proposition 1 and Theorem 1. Therefore, the convergence of our proposed dual state/parameter estimation filter to their optimal distributions, for $\{x_t,\theta_t\}\in D_{\mathcal{N}}$ as $N\rightarrow \infty$ can be investigated along the lines of Theorem 3.1 of \cite{shon}. This is left as a topic of our future research.
\end{remark}

\subsection{{Equivalent Flop (EF) Complexity Analysis of Dual State/Parameter Estimation Algorithm}}

{In this subsection, the computational complexity of our proposed dual state and parameter estimation algorithm is studied based on the equivalent flop (EF) complexity metric. This measure is introduced in \cite{flop}, in which the number of flops that result in the same computational time corresponding to a given operation is evaluated. The EF metric is mostly evaluated for those operations that depend on matrix and vector manipulations. The details regarding the EF complexity analysis for (a) our proposed dual estimation scheme, (b) the conventional Bayesian method for state and parameter estimation \cite{Parameter} (where Regularized particle filter structure is utilized to implement the filter), and (c) the recursive maximum likelihood (RML) method according to simultaneous perturbation stochastic approximation (SPSA) algorithm  \cite{poyi_gradientfree}   are presented in the Appendix C.}

{Let  $n_x,\;n_{\theta}.\;n_y$ denote the dimension of the state vector, the parameter vector, and the output vector, respectively, and $c_1$, $c_2$, $c_3$ denote the complexity of random number generation, resampling, and regularization, respectively. According to the summarized results in Tables \ref{table:state}, \ref{table:parameter}, \ref{table:baysian}, and \ref{table:rml} (in Appendix C), the EF complexity of the selected algorithms are given in Table \ref{table:compare}, where only the dominant parts of $C_{D}(n_x,n_{\theta},c_1,c_2,c_3,N)$ (this represents the EF complexity of our proposed dual estimation scheme), $C_{B}(n_x,n_{\theta},c_1,c_2,c_3,N)$ (this represents the EF complexity of the conventional Bayesian method \cite{Parameter}), and $C_{M}(n_x,n_{\theta},c_1,c_2,c_3,N)$ (this represents the recursive maximum likelihood (RML) method according to SPSA algorithm \cite{poyi_gradientfree})  are provided. This selection is justified by the fact that $N\gg 1$, therefore the dominant terms are related to $N$.}

{To achieve the \textit{same complexity} level in the three schemes, the number of the required particles in our proposed dual estimation scheme is determined based on the number of the particles that are utilized in the other two methods that are denoted by ${N}_{B}$ and ${N}_{M}$. Namely, we have
$N={N}_{B}(1-[2n_\theta^2+ 5n_{\theta}+2n_{\theta}n_y+6n_y+2n_{\theta}-6n_xn_{\theta}-c_3n_{\theta} ]/[3n_x^2+5n_\theta^2+6n_\theta+2n_\theta n_y+7n_y+3n_x+c_1(n_x+n_\theta)+c_2(n_x+n_\theta)+c_3n_x])$, and  
$N={N}_{M}(1-[n_x^2+5n_\theta^2+ 2n_{\theta}+n_x+2n_{\theta}n_y+7n_y+c_2n_{\theta} ]/[3n_x^2+5n_\theta^2+6n_\theta+2n_\theta n_y+7n_y+3n_x+c_1(n_x+n_\theta)+c_2(n_x+n_\theta)+c_3n_x])$}. 
{Considering the first equation 
we have} $2n_\theta^2+5n_{\theta}+2n_{\theta}n_y+6n_y+2n_{\theta}-6n_x n_{\theta}-c_3 n_{\theta}> 3n_x^2+5n_\theta^2+6n_\theta+2n_\theta n_y+7n_y+3n_x+c_1(n_x+n_\theta)+c_2(n_x+n_\theta)+c_3 n_x)$, {since the complexity corresponding to regularization ($c_3$) is assumed to be the dominant complexity term in the nominator of the coefficient of ${N}_{B}$. Therefore, the coefficient of ${N}_{B}$ is greater than that of $N$. Therefore, this  indicates that in order to achieve the same complexity by using the dual estimation algorithm and the one by using the conventional Bayesian method using regularized particle filter structure, the number of the particles required in the Bayesian method should be selected less than the number of particles in the dual estimation method. On the other hand, similar analysis for the RML method according to the second equation above  
shows that since} $n_x^2+5n_\theta^2+ 2n_{\theta}+n_x+2n_{\theta}n_y+7n_y+c_2n_{\theta} < 3n_x^2+5n_\theta^2+6n_\theta+2n_\theta n_y+7n_y+3n_x+c_1(n_x+n_\theta)+c_2(n_x+n_\theta)+c_3n_x)$, {unlike the Bayesian method, in the RML method one needs more particles in order to achieve the same computational complexity as that of the dual estimation scheme.

 Finally,  the EF complexity results are utilized to measure the time complexity of each algorithm considering the fact that the EF complexity is proportional to the time complexity of the algorithm.}
\begin{table*}
\caption{{The Approximated Total Equivalent Flop (EF) Complexity of the Selected \textit{Three Filters}.}}
\centering
\label{table:compare}
  \scalebox{0.85}{
   \centering
\begin{tabular}{|c|c|}
\hline
\multirow{1}{*}{Prediction Method} &  \multicolumn{1}{c|}{Total Equivalent Complexity}  \\ 
\hline
   Our Dual Estimation Scheme & $C_{B}(n_x,n_\theta ,c_1,c_2,c_3,N) \approx N(3n_x^2+5n_\theta^2+6n_\theta
   +2n_\theta n_y+7n_y+3n_x+c_1(n_x+n_\theta)+c_2(n_x+n_\theta)+c_3n_x) $  \\  \hline
Bayesian PF-based Estimation Method \cite{Parameter} &
  ${C}_{B}(n_x,n_\theta,c_1,c_2,c_3,N) \approx N(3n_x^2+3n_\theta^2+6n_xn_\theta+(1+c_1+c_2+c_3)n_x+(1+c_1+c_2+c_3)n_\theta+n_y) $  \\ \hline
RML  PF-based  Estimation Method  \cite{poyi_gradientfree}
  & ${C}_{M}(n_x,n_\theta,c_1,c_2,c_3,N) \approx N(2n_x^2+4n_\theta+2n_x+c_1(2n_x+n_\theta)+c_2 n_x+c_3 n_x) $  \\
   \hline
\end{tabular}
}
\end{table*}
{Our proposed dual state and parameter estimation scheme is an effective methodology for solving the fault diagnosis of nonlinear systems problem. Without loss of any generality one practically initiate operating the system from the healthy mode of operation. During the healthy operation of the system our proposed dual state and parameter estimation strategy can provide  an accurate and reliable information on the health parameters of the system. This information can then be readily used to perform the task of fault detection, isolation and identification of the system. This is due to the fact that  following the presence or injection of a fault in the components of the system, the deviation and changes in the health parameters do provide clear signals for the presence of a fault. In the next subsection, the application of our  approach developed in the previous subsections will be demonstrated and utilized for addressing the fault diagnosis problem of nonlinear systems.}
\vspace{-2mm}
\subsection{The Fault Diagnosis Formulation}
Determination and diagnosis of drifts in unmeasurable parameters of a system require an on-line parameter estimation scheme. In parametric modeling of a system anomaly or drift, generally it is assumed that the parameters are either constant or dependent on only the system states \cite{param1}. Hence, drifts in the parameters must be estimated through estimation techniques.

In \cite{isermann}, various parameter estimation techniques such as least squares, instrumental variables and estimation via discrete-time models have been surveyed. The main drawbacks and limitation with such methods arise due to the complexity and nonlinearity of the systems that we are considering in this paper that render the parameter estimation here a nonlinear optimization problem that must be solved in real-time. In \cite{2013} a nonlinear least squares (NLS) optimization scheme is developed for only the fault identification of a hybrid system.

Parameter estimation techniques that are used for fault diagnosis of system components generate residuals by comparing the estimated parameters that are obtained by either the ordinary least squares (OLS) or the recursive least squares (RLS) algorithms with parameters that are estimated under the initial fault free operation of the system \cite{simani}.

{The fault diagnosis problem under consideration  deals with simultaneously obtaining an optimal estimate of the system states as well as the \textit{time-varying}} parameters of a nonlinear system whose dynamics is governed by the discrete-time stochastic model,
\vspace{-2mm}
 \begin{eqnarray}\label{states1}
  x_{t+1}=f_t(x_{t},\theta_{t}^{\rm{T}}\lambda(x_{t}),\omega_{t}),
\end{eqnarray}
\vspace{-8mm}
\begin{eqnarray}\label{output1}
  y_{t}=h_t(x_{t},\theta_{t}^{\rm{T}}\lambda(x_{t}))+\nu_t,
 \vspace{-2mm}
\end{eqnarray}
where $f_{t}:\mathbb{R}^{n_{x}}\times \mathbb{R}^{n_\theta}\times \mathbb{R}^{n_\omega}\longrightarrow \mathbb{R}^{n_{x}}$ is a known  nonlinear function, $\theta_{t}\in \mathbb{R}^{n_{\theta}}$ is the unknown and  time-varying parameter vector that for a healthy system is set to ${\bf{1}}$,  $\lambda:\mathbb{R}^{n_{x}}\longrightarrow \mathbb{R}^{n_\theta}$ is a differentiable function that determines the relationship between the system states and the health parameters. The function $h_{t}:\mathbb{R}^{n_{x}}\times \mathbb{R}^{n_\theta}\longrightarrow \mathbb{R}^{n_{y}}$ is a known nonlinear function, $\omega_t$ and $\nu_t$ are uncorrelated noise sequences with covariance matrices $L_t$ and $V_t$, respectively. According to the formulation used in equations \eqref{states1} and \eqref{output1}, the parameter $\theta_t$ is a multiplicative fault vector whose value is considered to be set equal to ${\bf{1}}$ under the healthy mode of the system operation.

The model \eqref{states1} and \eqref{output1} is now used to investigate the problem of fault diagnosis (FD), which in this work is defined as the problem of \textit{fault detection, isolation, and identification} (FDII) when the system health parameters are considered to be affected by an \textit{unknown} and potentially time-varying multiplicative fault vector $\theta_t$.

The system health parameters are known functions of the system states, $\lambda(x_t)$, and the multiplicative fault vector $\theta_t$ is to be estimated. In other words, the {\it {a posteriori}} estimated parameter $\hat{\theta}_{t|t}$ will be used to generate residual signals for accomplishing the fault diagnosis goal and objective. {It is worth noting that based on our proposed formulation in equations \eqref{states1} and \eqref{output1} in order to capture variations of the system health parameters, these parameters are considered to be functions of the system states while their time variations due to degradations and anomalies are captured by introducing the fault vector. Therefore, changes due to variations in the system states are not considered as faults and determination of thresholds for developing the fault diagnosis decision making scheme will effectively be based on the healthy system in which the fault vector is represented and set to one.}

{The required residuals to determine the fault detection criterion  are obtained as the difference between the estimated parameters under the healthy operational mode that is denoted} by $\hat\theta_0$, and the estimated parameters under the faulty operational mode of the system that is denoted by $\hat \theta_{t|t}$ as follows
\vspace{-2mm}
\begin{equation}\label{threshold}
r_t=\hat{\theta}_0-\hat{\theta}_{t|t}.
\vspace{-1mm}
\end{equation}
It should be pointed out that the true value of the parameter is denoted by $\theta^\star _t$, which is assumed to be unknown.

For the implementation of our proposed fault diagnosis strategy that is constructed based on the previously developed state/parameter estimation framework, the parameter estimates will be considered as the main indicators for detecting, isolating, and identifying the faults in the system components. The residuals are generated from the parameter estimates under the healthy and faulty operational modes of the system according to equation \eqref{threshold}. The estimation of the parameters under the healthy operational mode is determined according to,
\vspace{-2mm}
\begin{eqnarray}\label{res}
\begin{split}
\hat\theta_0 = {\rm{arg max}}(-{\rm{log}}(\hat p(\theta_{0}|y_{1:T})),
  \end{split}
\vspace{-6mm}
 \end{eqnarray}
where $\hat{p}(\theta_0|y_{1:T})$ denotes the probability density  (conditioned on the observations up to time $T$ associated with the healthy data), that is obtained from the collected estimates and fitted to a normal distribution. The time window $T$ is chosen according to the convergence time of the parameter estimation algorithm. The thresholds to indicate the confidence intervals for each parameter are obtained through Monte Carlo analysis that is performed under different single-fault and multi-fault scenarios. The estimated parameters $\hat \theta_{t|t}$ are generated by following the procedure that was developed and proposed in  previous subsections.
\vspace{-4mm}
\section{Fault Diagnosis of a Gas Turbine Engine}\label{V}
In this section, the utility of our proposed dual estimation framework when applied to the problem of fault diagnosis of a nonlinear model of a gas turbine engine is demonstrated and investigated. The performance of our proposed state/parameter estimation scheme will be evaluated and investigated when the gas turbine is subjected to deficiencies in its health parameters due to injection of simultaneous faults.
\vspace{-4mm}
 \subsection{Model Overview}
The mathematical model of a gas turbine used in this paper is a single spool jet engine as developed in \cite{esmaeil}. The four engine states are the combustion chamber pressure and temperature, $P_{\rm{CC}}$ and $T_{\rm{CC}}$, respectively, the spool speed $S$, and the nozzle outlet pressure $P_{\rm{NLT}}$. The continuous-time state space model of the gas turbine is given as follows,
\vspace{-2mm}
{\small{
\begin{align}\label{system_eq}
\begin{split}
 &\dot{T}_{\rm{CC}}=\frac{1}{c_{\rm{v}} \dot m_{\rm{cc}}}[(c_{\rm{p}}T_{\rm{C}}\theta_{m_{\rm{C}}}\dot{m}_{\rm{C}}+\eta_{\rm{CC}}H_{\rm{u}}\dot{m}_{\rm{f}}
 -c_{\rm{p}}T_{\rm{CC}}\theta_{m_{\rm{T}}}\dot{m}_{\rm{T}}) -\\&c_{\rm{v}}T_{\rm{CC}}(\theta_{m_{\rm{C}}}\dot{m}_{\rm{C}}+\dot{m}_{\rm{f}}-\theta_{m_{\rm{T}}}\dot{m}_{\rm{T}})],\\
 &\dot{S}=\frac{\eta_{\rm{mech}} \theta_{m_{\rm{T}}}\dot{m}_{\rm{T}} c_{\rm{p}} (T_{\rm{CC}}-T_{\rm{T}}) - \theta_{m_{\rm{C}}}\dot{m}_{\rm{C}} c_{\rm{p}} (T_{\rm{C}}-T_{\rm{d}})}{JS(\frac{\pi}{30})^{2}},  \\
 &\dot{P}_{\rm{CC}}=\frac{P_{\rm{CC}}}{T_{\rm{CC}}}\frac{1}{c_{\rm{v}}\dot m_{\rm{cc}}}[(c_{\rm{p}}T_{\rm{C}}\theta_{m_{\rm{C}}}\dot{m}_{\rm{C}}+\eta_{\rm{CC}}H_{\rm{u}}\dot{m}_{\rm{f}}
 -c_{\rm{p}}T_{\rm{CC}}\theta_{m_{\rm{T}}}\dot{m}_{\rm{T}})-\\&c_{\rm{v}}T_{\rm{CC}}(\theta_{m_{\rm{C}}}\dot{m}_{\rm{C}}+\dot{m}_{\rm{f}}-\theta_{m_{\rm{T}}}\dot{m}_{\rm{T}})]
+\frac{\gamma{\rm{R}}{T_{\rm{CC}}}}{V_{\rm{CC}}}(\theta_{m_{\rm{C}}}\dot{m}_{\rm{C}}+\dot{m}_{\rm{f}}-\theta_{m_{\rm{T}}}\dot{m}_{\rm{T}}),\\
 &\dot{P}_{\rm{NLT}}=\frac{{T_{\rm{M}}}}{V_{\rm{M}}}(\theta_{m_{\rm{T}}}\dot{m}_{\rm{T}}+\frac{\beta}{\beta+1}\theta_{m_{\rm{C}}}\dot{m}_{\rm{C}}-\dot{m}_{\rm{Nozzle}}).
 \end{split}
\end{align}}}
For the physical significance of the model parameters and details refer to \cite{esmaeil}. The five gas turbine measured outputs are considered to be the compressor temperature ($y_1$), the combustion chamber pressure ($y_2$), the spool speed ($y_3$), the nozzle outlet pressure ($y_4$), and the turbine temperature ($y_5$), namely
\begin{eqnarray}\label{out_eq}
\begin{split}
y_{1}&=T_{\rm{C}}=T_{\rm{diffuser}}[1+\frac{1}{\theta_{\eta_{\rm{C}}}\eta_{\rm{C}}}[(\frac{P_{\rm{CC}}}{P_{\rm{diffuzer}}})^{\frac{\gamma-1}{\gamma}}-1]],\\
y_{2}&=P_{\rm{CC}},\ y_{3}=S,\ y_{4}=P_{\rm{NLT}},\\
 y_{5}&=T_{\rm{CC}}[1-\theta_{\eta_{\rm{T}}}\eta_{\rm{T}}(1-(\frac{P_{\rm{NLT}}}{P_{\rm{CC}}})^{\frac{\gamma-1}{\gamma}}].
\end{split}
\end{eqnarray}
In order to discretize the above model for implementation of our proposed dual state/parameter estimation particle filters, a simple Euler Backward method is applied with the sampling period of $T_s= 10\;{\rm{msec}}$.

The system health parameters are represented by the compressor and the turbine efficiency, $\eta_{\rm{C}}$ and $\eta_{\rm{T}}$, respectively, and the compressor and turbine mass flow capacities, $\dot m_{\rm{C}}$ and $\dot m_{\rm{T}}$, respectively. A fault vector is incorporated in the above model to manifest the effects of system health parameters that are denoted by $\theta=[\theta_{\eta_{\rm{C}}},\theta_{m_{\rm{C}}},\theta_{\eta_{\rm{T}}},\theta_{m_{\rm{T}}}]^{\rm{T}}$. By introducing a new parameter as $\acute {\theta}_{\eta_{\rm{C}}}=\frac{1}{\theta_{\eta_{\rm{C}}}}$, the measurement equations \eqref{out_eq} can be represented as smooth functions with respect to the fault parameters. Each parameter variation can be a manifestation of changes in the fault vector that is considered as a multiplicative fault type. All the simulations that are conducted in this section corresponds to the cruise flight condition mode.

{In order to demonstrate the effectiveness and capabilities of our proposed algorithms, we have also conducted  simulation results corresponding to \textit{two} other techniques in the literature, namely (a) the conventional Bayesian method \cite{Parameter}, and (b) the well-known recursive maximum likelihood (RML) parameter estimation method based on PF \cite{poyi_2005, poyi_2011, poyi_gradientfree}. It should be noted that the number of particles in each algorithm is chosen based on the execution time of the algorithm such that approximately the same execution time is achieved for each algorithm. Moreover, a \textit{third} method known as the gradient free PF-based RML method \cite{poyi_gradientfree} could not  yield  an acceptable performance in this application given the large number of the tuning parameters that are required and associated with each parameter in this method. Therefore, the RML based on the direct gradient method is utilized for the purpose of our performance comparison analysis.}

In our schemes, the adaptive step size ($P_t^{(j)}=\gamma_tR_t^{(j)}$) is defined as the product of the constant $\gamma_t (\gamma_t=0.9)$ with $R_t^{(j)}$, which is evaluated on-line from the trace of the prediction error covariance matrix that is estimated from the maximum likelihood method. On the other hand, the step size in the RML method was chosen as $\gamma_t =0.05={\rm{const.}}$ that is obtained by trial and error. The residuals corresponding to the parameter estimates are also obtained. Based on the percentage of the maximum absolute error criterion, a convergence time of 2 seconds is obtained  for estimating both the states and parameters corresponding to 25 Monte Carlo runs of simultaneous faults with severities ranging from $1\%$ to $10\%$ loss of effectiveness of the healthy condition.

To choose the number of particles for implementation of the state and parameter estimation filters, a quantitative study {is now conducted. Specifically, based on the mean absolute error (MAE$\%$) that was obtained at the estimation process steady state  and by taking into account the algorithm's computational time, the number of particles is chosen as $N=50$ for both the state and parameter estimation filters in this application. On the other hand, by taking into account the average execution time of $18 sec$ for one iteration of the dual estimation algorithm, an equivalent execution time can be achieved for the Bayesian algorithm with ${N}_B=45$ and for the RML algorithm with ${N}_M=150$ particles.}

{We have confirmed through simulations that acceptable performance and convergence times are obtained. The shrinkage matrix is selected as $0.93I$. The initial distributions (i.e., the mean and covariance matrices) of the states and parameters are selected to correspond to the cruise flight operational condition as provided in \cite{esmaeil}. In what follows, the two main case scenarios for performing fault diagnosis  of the gas turbine engine are presented.}

{\it{\underline {\textbf{Scenario I: Concurrent Faults in the Compressor and Turbine Health Parameters.}}}}

In this scenario, the input fuel flow rate to the engine is changed by decreasing it by $2\%$ from its nominal value one second after reaching the steady state condition. Next, the effects of concurrent faults in both the compressor and the turbine are studied by injecting sequential fault patterns affecting the system components. Specifically, at time $t=4\;{\rm{sec}}$ the compressor efficiency is reduced by $5\%$ (this represents the level of the fault severity), followed by at $t=9\;{\rm{sec}}$ the same fault type affecting the compressor mass flow capacity, and at $t=14\;{\rm{sec}}$ the same fault type affecting the turbine efficiency, and finally at $t=19\;{\rm{sec}}$ the same fault type is applied to the turbine mass flow capacity.

{The results corresponding to changes in the fault parameters are depicted in Figure \ref{fig:res_con}. The dotted lines depict the confidence bounds for residuals that are determined based on $50$ Monte Carlo simulation runs under various concurrent and simultaneous single and/or multiple fault scenarios by using the PE-based method. By analyzing the residuals, the detection time of a fault in each component and its severity can be determined and identified. It follows from this figure that the  residuals corresponding to the dual estimation method using PE-based method \textit{as well as} the RML method practically \textit{do not} exceed their confidence bounds subject to changes in the engine input (which is applied at $t=1\;{\rm{sec}}$). On the other hand, the Bayesian method shows false alarms in the residuals corresponding to the turbine heath parameters. Moreover,  this method is also not capable of tracking the changes in the fault vector in the selected time window in all the residual signals after fault occurrence.}
 \begin{figure*}[htbp]
 \vspace{-3mm}
\hspace*{-2cm}
\includegraphics[scale=0.3]{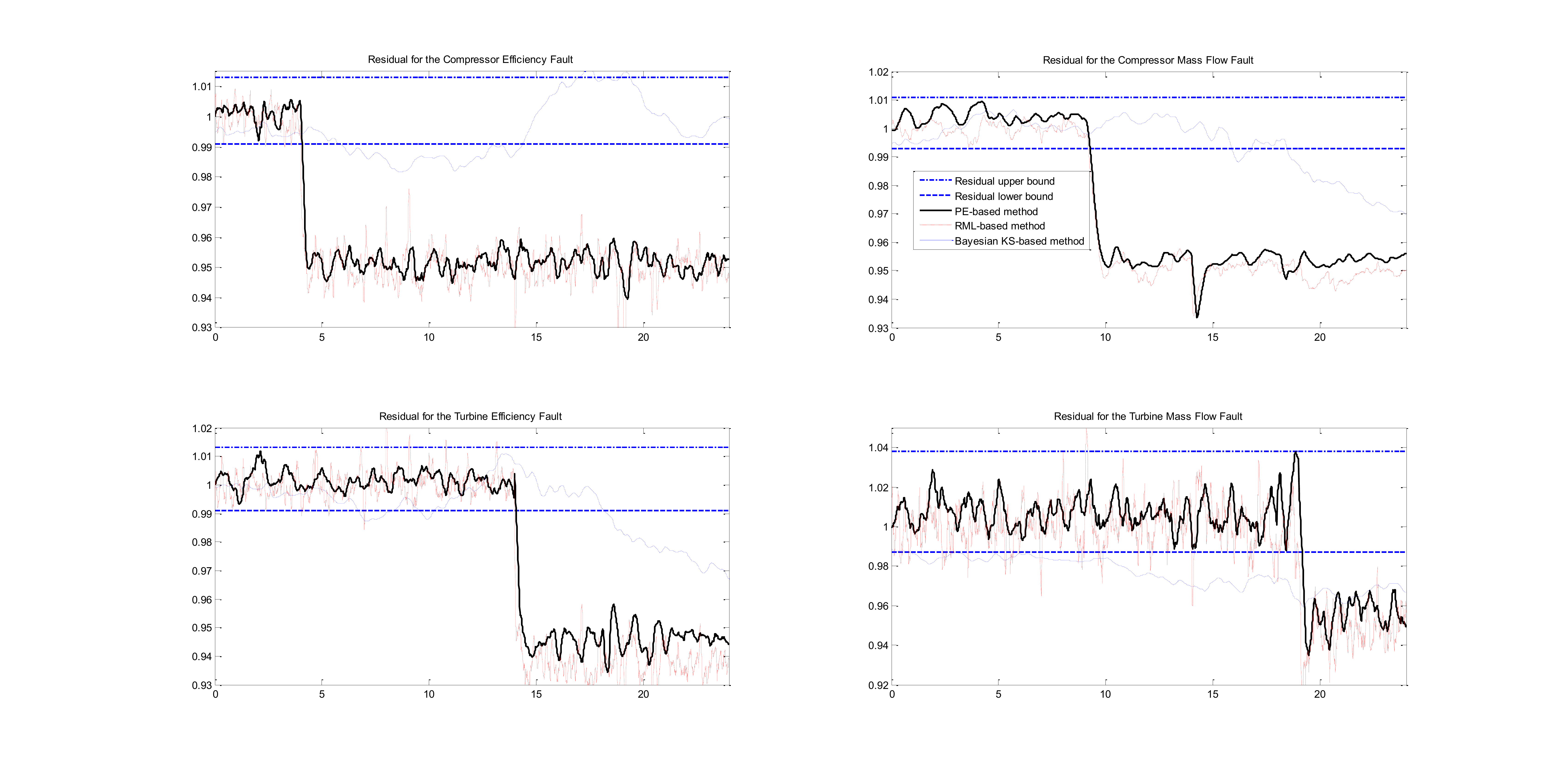}
 \vspace{-5mm}
\caption{{Residuals corresponding to the concurrent fault scenarios in the turbine and the compressor parameters.}}
\label{fig:res_con}
\vspace{-3mm}
\end{figure*}

In order to obtain a quantitative measure on the accuracy and reliability of our proposed estimation algorithm the results related to the $5\%$ fault severity using the mean absolute error (MAE$\%$) metric of estimates corresponding to the last $2\;{\rm{sec}}$ of simulations (following the algorithm convergence) after each change are provided and summarized in Table \ref{table:con_states}. {The state/parameter estimation MAE$\%$ for (a) our proposed dual estimation algorithm according to the PE-based method with $N=50$, (b) the conventional Bayesian method based kernel smoothing (KS-based) with ${N}_{B}=45$, and (c) the RML method with ${N}_{M}=150$ are presented. In this table, the $i$-th fault for $i=1,...,4$ denotes the last $2\;{\rm{sec}}$ of simulations after the $i$-th fault occurrence, and the first column refers to the healthy system before the fault occurrence.}

The results shown in Table \ref{table:con_states} demonstrate that for the PE-based method the maximum MAE$\%$ for the states is between $0.03\%-1.06\%$ of their nominal values. In case of the health parameters, for $ \eta_{\rm{C}}$ and $\dot m_{\rm{C}}$ the maximum MAE$\%$ is around $0.91\%$ and for $ \eta_{\rm{T}}$ and $\dot m_{\rm{T}}$ it is around $0.98\%$ of their nominal values. On the other hand, according to the results presented in Table \ref{table:con_states} (b), the maximum MAE$\%$ for the states corresponding to the RML method is between $0.1\%-1.16\%$ of their nominal values. In case of the health parameters, the maximum MAE$\%$ ranges between $0.8\%-2.8\%$ of their nominal values, where both mass flow rates are estimated with higher MAE$\%$. { The results corresponding to Table \ref{table:con_states} (c) indicate that the maximum MAE$\%$ for the state estimation results in the Bayesian KS-based method ranges between $0.2\%-18.3\%$, for the compressor health parameters between $0.53\%-19.3\%$, and for the turbine health parameters between $0.15\%-3.0\%$.}

 The MAE$\%$ for the estimated measurements (outputs) of the engine are also provided in Table \ref{table:con_outputs} for the PE-based, the RML, and the Bayesian KS-based methods. From the results presented in Table \ref{table:con_outputs} (a) one can conclude that the maximum MAE$\%$ for the temperature (of the turbine and the compressor) corresponding to our proposed PE-based method is less than $0.3\%$, and for the spool speed it is less than $0.16\%$, and for the compressor pressure it is less than $1.4\%$, and for the turbine pressure it is less than $2.5\%$. On the other hand, the results presented in Table \ref{table:con_outputs} (b) for the RML method show that the maximum MAE$\%$ for the compressor and turbine temperatures is less than $0.4\%$ and $0.6\%$, respectively. For the spool speed the MAE$\%$ is less than $0.2\%$, and for the compressor pressure it is less than $1.5\%$ and for the turbine pressure it is less than $2.5\%$. {For the Bayesian KS-based method, instead of the compressor temperature and spool speed outputs, the maximum MAE$\%$ exceeds $13\%$ of the nominal values. Consequently, the results presented in the above two tables confirm that the  Bayesian KS-based method \textit{does not} have acceptable estimation accuracy as compared to the other two alternative methods. On the other hand, the PE-based method outperforms the RML method significantly. The accuracy in the measurement estimation is an important consideration  and factor given that from practical perspectives the system states and parameters are unknown.} Therefore, it is generally necessary to judge the estimation accuracy based on the output estimation error performance and behavior.

In order to demonstrate and illustrate the accuracy of our proposed fault detection algorithm based on the dual state/parameter estimation scheme, at the end of this section a quantitative study is conducted by performing a {\textit{confusion matrix analysis}} \cite{confusion_2} in presence of various fault scenarios having different fault severities and in presence of the same level of process and measurement noise that are stated in \cite{esmaeil} for the PE-based method, the RML method, and the Bayesian KS-based combined state and parameter estimation algorithm.
\begin{table}
\caption{{State/Parameter MAE$\%$ in case of concurrent fault scenarios for (a) The Dual estimation algorithm according to PE-based method with $N=50$, (b) The RML method with $N=150$, and (c) The Bayesian KS-based method with $N=45$.}}
\label{table:con_states}
\vspace{-5mm}
\subtable[]{
\centering
\begin{tabular}{|c|c|c|c|c|c|c|c|}
\hline
State &$\  {\rm{No\; Fault}}$ & $1nd \  {\rm{Fault}}$& $2rd  \ {\rm{Fault}}$& $3th \ {\rm{Fault}}$& $4th \  {\rm{Fault}}$  \\
\hline
\hline                  
   $P_{\rm{CC}}$ &0.3529    &0.2097    &0.3614    &0.4336    &0.2374\\
  $N$  &0.1473    &0.0761    &0.1087    &0.1624   & 0.0296    \\
  $T_{\rm{CC}}$& 0.2683   & 0.1674   & 0.1678    &0.3838    &0.1155\\
   $P_{\rm{NLT}}$ & 0.8575    &0.5325    &0.3978   & 1.0614    &0.3213 \\
   $ \eta_{\rm{C}}$ &  0.2702   & 0.1785    &0.2749    &0.3879   & 0.2107\\
   $\dot m_{\rm{C}}$ &0.6621   &0.4229    &0.3682   & 0.9132    &0.2236\\
   $ \eta_{\rm{T}}$  &0.2865    &0.1648    &0.1743    &0.4885    &0.1873\\
   $\dot m_{\rm{T}}$ & 0.4744   & 0.4557    &0.4889   & 0.9757   & 0.5037\\
\hline
\end{tabular}
}\\
\subtable[]{
\centering
\begin{tabular}{|c|c|c|c|c|c|c|c|}
\hline
State &$\  {\rm{No\; Fault}}$ & $1nd \  {\rm{Fault}}$& $2rd  \ {\rm{Fault}}$& $3th \ {\rm{Fault}}$& $4th \  {\rm{Fault}}$  \\
\hline
\hline                  
   $P_{\rm{CC}}$  &0.5352    &0.6342    &0.3921    &0.8934    &0.5882\\
  $N$  & 0.0995    &0.0912    &0.1018    &0.2060    &0.1383    \\
  $T_{\rm{CC}}$&  0.2064    &0.2443    &0.2574    &0.5174    &0.3374\\
   $P_{\rm{NLT}}$ &  0.7181    &0.8112    &0.7771    &1.1603    &0.5666 \\
   $ \eta_{\rm{C}}$ &  0.9268    &1.9195    &2.1698    &1.4508    &1.4651\\
   $\dot m_{\rm{C}}$ &1.6338    &1.8037    &1.0761   &2.6062   &2.2717\\
   $ \eta_{\rm{T}}$  & 0.9252    &0.7876    &0.8714    &1.6517    &1.1411\\
   $\dot m_{\rm{T}}$ & 1.3719    &1.7858    &1.6162    &1.9666    &2.7653\\
\hline
\end{tabular}
}\\
\subtable[]{
\centering
\begin{tabular}{|c|c|c|c|c|c|c|c|}
\hline
State &$\  {\rm{No\; Fault}}$ & $1nd \  {\rm{Fault}}$& $2rd  \ {\rm{Fault}}$& $3th \ {\rm{Fault}}$& $4th \  {\rm{Fault}}$  \\
\hline
\hline                  
   $P_{\rm{CC}}$  & 1.8961    &2.6032    &6.1590  & 18.2816    &7.9636\\
  $N$  &  0.2127    &0.5032    &0.4490    &4.7275    &2.5564\\
  $T_{\rm{CC}}$&   0.4789    &1.0029    &1.6025    &8.6930    &6.5029\\
   $P_{\rm{NLT}}$ &  0.6841    &1.3838   & 3.6558   &14.9288   & 8.9250\\
   $ \eta_{\rm{C}}$ &  0.7248    &3.3660    &3.0788    &6.3141   & 4.9584\\
   $\dot m_{\rm{C}}$ & 0.5306    &1.3399    &4.3086   &19.3026   &11.8476\\
   $ \eta_{\rm{T}}$  &  0.1445    &0.7394   & 0.9198    &1.0082    &1.8379\\
   $\dot m_{\rm{T}}$ &  1.4943    &1.6979    &2.6633   & 3.0198   & 1.7125\\
\hline
\end{tabular}
}
\end{table}
\begin{table}
\caption{{Output estimation MAE$\%$ in case of concurrent fault scenarios for (a) The Dual estimation algorithm according to PE-based method with $N=50$, (b) The RML method with $N=150$, and  (c) The Bayesian KS-based method with $N=45$.}}
\label{table:con_outputs}
\vspace{-5mm}
\subtable[]{
\centering
\begin{tabular}{|c|c|c|c|c|c|c|c|}
\hline
Output &$\  {\rm{No\; Fault}}$ & $1nd \  {\rm{Fault}}$& $2rd  \ {\rm{Fault}}$& $3th \ {\rm{Fault}}$& $4th \  {\rm{Fault}}$  \\
\hline
\hline                  
   $T_{\rm{C}}$ &0.2893   & 0.2319   & 0.2749    &0.2805   & 0.2357 \\
   $P_{\rm{C}}$& 1.3548    &1.2332   & 1.2507   & 1.4070    &1.1813\\
   $N$ &  0.1473   & 0.0761   & 0.1087   & 0.1624    &0.0296\\
   $T_{\rm{T}}$ &  0.2034  &  0.1857   & 0.1911  &  0.2804    &0.1322\\
   $P_{\rm{T}}$ &2.2231   & 2.1696   & 2.1839   & 2.4577   & 2.0783\\
\hline
\end{tabular}
}\\
\subtable[]{
\begin{tabular}{|c|c|c|c|c|c|c|c|}
\hline
Output &$\  {\rm{No\; Fault}}$ & $1nd \  {\rm{Fault}}$& $2rd  \ {\rm{Fault}}$& $3th \ {\rm{Fault}}$& $4th \  {\rm{Fault}}$  \\
\hline
\hline                  
   $T_{\rm{C}}$ &0.2902    &0.3240    &0.2956    &0.3985    &0.2991 \\
   $P_{\rm{C}}$& 1.4012    &1.3755    &1.3030    &1.4779    &1.2902\\
   $N$ &  0.0995    &0.0912    &0.1018    &0.2060   & 0.1383\\
   $T_{\rm{T}}$ &  0.2181   & 0.2461   & 0.2122    &0.5786    &0.5206\\
   $P_{\rm{T}}$ &2.3446    &2.3994    &2.1356    &2.5220    &2.2474\\
\hline
\end{tabular}
}\\
\subtable[]{
\begin{tabular}{|c|c|c|c|c|c|c|c|}
\hline
Output &$\  {\rm{No\; Fault}}$ & $1nd \  {\rm{Fault}}$& $2rd  \ {\rm{Fault}}$& $3th \ {\rm{Fault}}$& $4th \  {\rm{Fault}}$  \\
\hline
\hline                  
   $T_{\rm{C}}$ &0.9416    &0.7569    &0.6033    &3.4434   & 0.3789\\
   $P_{\rm{C}}$&  1.9784    &2.6211    &6.5484   &18.0905    &7.8644\\
   $N$ &   0.2127    &0.5032    &0.4490    &4.7275    &2.5564\\
   $T_{\rm{T}}$ & 0.3428   & 0.7125   & 2.6660  & 13.3328    &8.6219\\
   $P_{\rm{T}}$ & 2.2715   & 2.7023   & 3.9467  & 14.9107    &8.7964\\
\hline
\end{tabular}
}
\end{table}

{\it{\underline{\textbf{Scenario II: Simultaneous Faults in the Compressor and the Turbine Health Parameters.}}}}

In the second scenario, a simultaneous fault in \textit{all} the 4 health parameters of the engine is applied at $t=9\;{\rm{sec}}$. The compressor and the turbine efficiencies faults follow the pattern of a drift fault that starts at $t=9\;{\rm{sec}}$ and causes a $5\%$ loss of effectiveness in the compressor efficiency by the end of the simulation time (i.e. at $t=19\;{\rm{sec}}$), and a $3\%$ loss of effectiveness in the turbine efficiency by the end of the simulation time. Simultaneously, the mass flow rate capacities of both the compressor and the turbine are affected by a  $5\%$ loss of effectiveness fault.

{The residuals corresponding to the three previous estimation methods are provided in Figure \ref{fig:res_sim}. The simulations show that in case of changes in the engine input (applied at $t=1\;{\rm{sec}}$) the RML method residuals has high false alarm rates as compared to dual estimation method according to PE-based algorithm, similar to the first scenario for the concurrent faults. More quantitative analysis on the performance of the RML method that is compared to the PE-based method is provided in the subsequent subsection. The presented results admit that the Bayesian KS-based method is not able to track the variations in the fault vectors in the case of simultaneous fault scenario.}

The results in Table \ref{table:sim_states} (a) show that for our proposed PE-based method, the maximum MAE$\%$ for both state and parameter estimates are between $0.1\%-0.5\%$ of their nominal values. However, in the worst case the post fault estimated MAE$\%$ of the $\dot m_{\rm{T}}$ is $0.47\%$ of its nominal value. Moreover, in the results shown for the RML method in Table \ref{table:sim_states} (b) it follows clearly that the state estimation MAE$\%$ can be achieved within $0.1\%-0.8\%$ of the nominal values, whereas the parameter estimation MAE$\%$ is achieved within $0.7\%-3\%$ of the nominal values with higher error rates after the fault occurrence, specially in the compressor and turbine mass flow rate capacities. {However, for the Bayesian KS-based method in Table \ref{table:sim_states} (c) the maximum MAE$\%$ is achieved within $0.19\%-8.4\%$ of the nominal values for the estimated states and within $0.25\%-7.2\%$ of the nominal values for the estimated parameters.}

 The MAE$\%$ measurement (output) estimate error given in Table \ref{table:sim_outputs} (a) for the PE-based method shows that after simultaneous fault occurrences the error increases when compared to their values before the fault occurrences. This is caused due to accumulation of parameter estimation errors while all the four parameters are affected by a fault. {On the other hand, the results corresponding to the output estimation as given in Tables \ref{table:sim_outputs}(a)-(c) show that with the exception of the turbine pressure, our PE-based method outperforms the RML method for estimating the other four measurement outputs. However, the maximum MAE$\%$ for the outputs from Bayesian KS-based method performs high level of errors after fault occurrence in all measurement outputs as compared to the other two estimation methods.}

To summarize, our proposed PE-based fault diagnosis algorithm is capable of detecting, isolating and estimating the component faults of a gas turbine engine with an average accuracy of $0.3\%$ for the compressor and $0.5\%$ for the turbine faults. In contrast the RML algorithm is capable of achieving the performance of an average $3\%$ for the compressor and $1.6\%$ for the turbine faults. {The Bayesian KS-based method does not have acceptable accuracy for simultaneous fault diagnosis application.}
\begin{figure*}[htbp]
 \vspace{-3mm}
\hspace*{-2cm}
\includegraphics[scale=0.3]{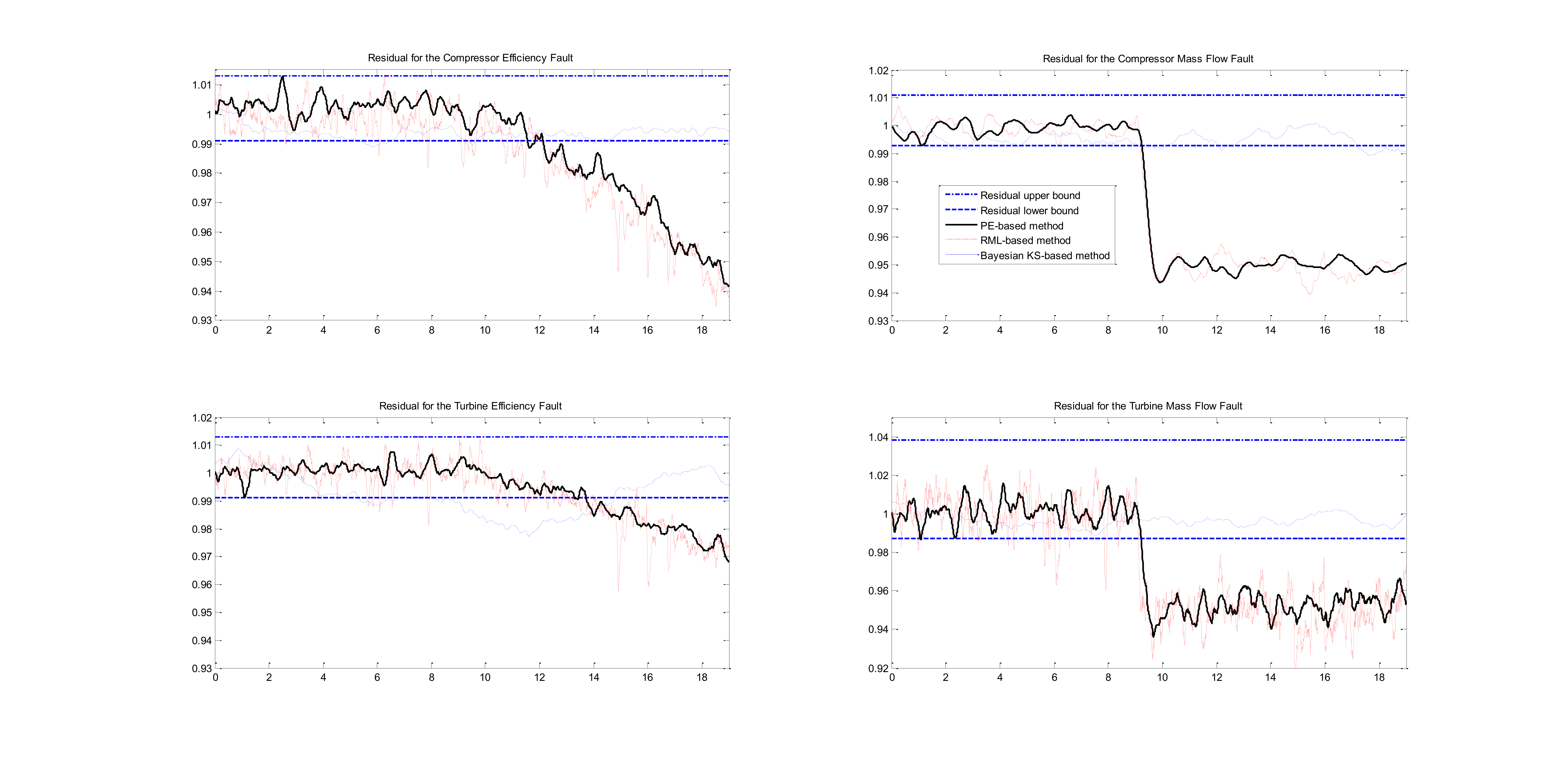}
 \vspace{-5mm}
\caption{{Residuals corresponding to the simultaneous fault scenarios.}}
\label{fig:res_sim}
\vspace{-3mm}
\end{figure*}
\begin{table}
\caption{{State/Parameter MAE$\%$ in case of simultaneous fault scenarios for (a) The PE-based method with $N=50$,  (b) the RML method with $N=150$, and (c) the Bayesian KS-based method with $N=45$.}}
\label{table:sim_states}
 \vspace{-5mm}
\centering
\subtable[]{
\footnotesize
\centering
\begin{tabular}{|c|c|c|}
\hline
State &$ {\rm{Before \  Fault}}$ & ${\rm{After \  Fault}}$  \\
\hline
\hline
   $P_{\rm{CC}}$ & 0.2217  &  0.2372 \\
  $N$  &0.0535   &0.1061 \\
  $T_{\rm{CC}}$& 0.2086   & 0.1928  \\
   $P_{\rm{NLT}}$ & 0.3970    & 0.4291 \\
   $ \eta_{\rm{C}}$ & 0.1735  &  0.1821 \\
   $\dot m_{\rm{C}}$ &0.2811  &  0.3293  \\
   $ \eta_{\rm{T}}$  &  0.1016  &  0.1485  \\
   $\dot m_{\rm{T}}$ &  0.4589  &  0.4744\\
\hline
\end{tabular}
}
\subtable[]{
\footnotesize
\centering
\begin{tabular}{|c|c|c|}
\hline
State &$ {\rm{Before \  Fault}}$ & ${\rm{After \  Fault}}$  \\
\hline
\hline
   $P_{\rm{CC}}$ & 0.4865  &  0.4383 \\
  $N$  &0.1025   &0.1053 \\
  $T_{\rm{CC}}$& 0.2247   &  0.1888  \\
   $P_{\rm{NLT}}$ & 0.7540    &  0.5264 \\
   $ \eta_{\rm{C}}$ & 0.9295  &  1.7956 \\
   $\dot m_{\rm{C}}$ &1.7291  &  2.9729  \\
   $ \eta_{\rm{T}}$  & 0.7306 &  1.0040  \\
   $\dot m_{\rm{T}}$ &  1.4290  &   1.5923\\
\hline
\end{tabular}
}
\subtable[]{
\footnotesize
\centering
\begin{tabular}{|c|c|c|}
\hline
State &$ {\rm{Before \  Fault}}$ & ${\rm{After \  Fault}}$  \\
\hline
\hline
   $P_{\rm{CC}}$ & 0.5186  &8.3590 \\
  $N$  & 0.1906   &2.7197    \\
  $T_{\rm{CC}}$&  0.2834   &4.5219 \\
   $P_{\rm{NLT}}$ &  0.6506   &5.0157 \\
   $ \eta_{\rm{C}}$ & 0.5992    &4.8358\\
   $\dot m_{\rm{C}}$ & 0.5775   &7.2215  \\
   $ \eta_{\rm{T}}$  &  0.2463    &4.8958\\
   $\dot m_{\rm{T}}$ &  0.2861   &4.3080\\
\hline
\end{tabular}
}
\vspace{-4mm}
\end{table}
\begin{table}
\caption{{Output estimation MAE$\%$ in case of simultaneous fault scenarios for (a) The PE-based method with $N=50$, (b) The RML method with $N=150$, and (c) The Bayesian KS-based method with $N=45$.}}
\label{table:sim_outputs}
 \vspace{-5mm}
\centering
\subtable[]{
\small
\centering
\begin{tabular}{|c|c|c|c|c|c|c|c|}
\hline
Output &$ {\rm{Before \  Fault}}$ & ${\rm{After \  Fault}}$  \\
\hline
\hline
   $T_{\rm{C}}$ &0.2207  &  0.2852   \\
   $P_{\rm{C}}$& 1.2926   &  1.3729 \\
   $N$ & 0.0535 &  0.1027  \\
   $T_{\rm{T}}$ &  0.1565 &   0.1650 \\
   $P_{\rm{T}}$ &2.1210 &   2.2348 \\
\hline
\end{tabular}
}
\subtable[]{
\small
\centering
\begin{tabular}{|c|c|c|c|c|c|c|c|}
\hline
Output &$ {\rm{Before \  Fault}}$ & ${\rm{After \  Fault}}$  \\
\hline
\hline
   $T_{\rm{C}}$ & 0.2482  &   0.2832   \\
   $P_{\rm{C}}$& 1.4051   &  1.4369 \\
   $N$ & 0.1025 &  0.1053  \\
   $T_{\rm{T}}$ &   0.2058 &   0.1774 \\
   $P_{\rm{T}}$ &2.1413 &    2.1580 \\
\hline
\end{tabular}
}
\subtable[]{
\small
\centering
\begin{tabular}{|c|c|c|c|c|c|c|c|}
\hline
Output &$ {\rm{Before \  Fault}}$ & ${\rm{After \  Fault}}$  \\
\hline
\hline
   $T_{\rm{C}}$ &  0.3250  & 3.5756   \\
   $P_{\rm{C}}$&  1.3621    &8.1037 \\
   $N$ & 0.1906   & 2.7197    \\
   $T_{\rm{T}}$ &  0.4126  & 4.9311 \\
   $P_{\rm{T}}$ & 2.2252   &5.0404\\
\hline
\end{tabular}
}
\vspace{-4mm}
\end{table}
\subsection*{\underline{Fault Diagnosis Confusion Matrix Analysis}}
Finally, in this subsection a quantitative study is performed by utilizing the confusion matrix analysis \cite{confusion_2} to evaluate the increase in the false alarms and/or misclassification rates of the faults in our considered application when the fault diagnosis algorithm is implemented by our proposed PE-based method with $N=50$ particles, the RML method with $N=150$, and the Bayesian KS-based method with $N=45$ particles. The thresholds corresponding to each algorithm are determined from $25$ Monte Carlo runs on simultaneous fault scenarios that are not necessarily the same for the three algorithms. The confusion matrix data is obtained by performing simulations for another $35$ Monte Carlo simultaneous fault scenarios having different fault severities and in presence of the same process and measurement noise covariances corresponding to $50\%$ of the nominal values of the process and measurement noise covariances (according to \cite{esmaeil}). In these scenarios, at each time more than one of the system health parameters are affected by component faults.

{The results are shown in Tables \ref{table:conf_50}-\ref{table:conf_RPE} corresponding to PE-based method with $N=50$ particles, the RML method with $N=150$, the RML method with $N=150$, and the Bayesian KS-based method with $N=45$ particles, respectively}. In these tables the rows depict the actual number of fault categories that are applied and the columns represent the number of estimated fault categories. The diagonal elements represent the true positive rate ($TP$) for each fault occurrence.
\begin{table}
\caption{{Confusion matrix for (a) The PE-based method with $N=50$, (b) The RML method with $N=150$, and (c) The Bayesian KS-based method with $N=45$.}}
\vspace{-3mm}
\centering
\subtable[]{
\small
\centering
\begin{tabular}{|c|c|c|c|c|c|c|c|}
\hline
Fault &$\dot \eta_{\rm{C}}$& $\dot m_{\rm{C}}$& $\dot \eta_{\rm{T}}$& $\dot m_{\rm{T}}$& $\rm{No\; Fault}$  \\
\hline
\hline                  
   $ \eta_{\rm{C}}$ &{\bf{31}}  &  0   & 2  &  2  &  0 \\
  $\dot m_{\rm{C}}$  &0&  {\bf{30}}  &  2 &  3  & 0    \\
  $ \eta_{\rm{T}}$& 1  &  1   & {\bf{28}}  &  4  &  1\\
   $\dot m_{\rm{T}}$ & 1 &  1  &  3 &   {\bf{29}} &  1 \\
   $ \rm{No\; Fault}$ &  0 &   0 &  1 &   1 &   {\bf{33}}\\
\hline
\end{tabular}
\label{table:conf_RPE}
}
\subtable[]{
\small
\centering
\begin{tabular}{|c|c|c|c|c|c|c|c|}
\hline
Fault &$\dot \eta_{\rm{C}}$& $\dot m_{\rm{C}}$& $\dot \eta_{\rm{T}}$& $ \dot m_{\rm{T}}$& $\rm{No\; Fault}$  \\
\hline
\hline                  
   $\eta_{\rm{C}}$ &{\bf{28}}  &  2   & 3  &  2  &  0 \\
  $\dot m_{\rm{C}}$  &1&  {\bf{27}}  &  1 &  4  & 2    \\
  $ \eta_{\rm{T}}$& 2  &  3   & {\bf{26}}  &  3  &  1\\
   $\dot m_{\rm{T}}$ & 1 &  3  &  4 &   {\bf{26}} &  1 \\
   $\rm{No\; Fault}$ &  0 &   2 &  1 &   1 &   {\bf{31}}\\
\hline
\end{tabular}
\label{table:conf_150}
}
\subtable[]{
\small
\centering
\begin{tabular}{|c|c|c|c|c|c|c|c|}
\hline
Fault &$\dot \eta_{\rm{C}}$& $\dot m_{\rm{C}}$& $\dot \eta_{\rm{T}}$& $\dot m_{\rm{T}}$& $\rm{No\; Fault}$  \\
\hline
\hline                  
   $ \eta_{\rm{C}}$ &{\bf{10}}  &  5   & 6  &  4  &  {\bf{10}} \\
  $\dot m_{\rm{C}}$  &9&  {\bf{13}}  &  8 &  6  & 9    \\
  $ \eta_{\rm{T}}$& 6  &  6   & {\bf{9}}  &  7 &  7\\
   $\dot m_{\rm{T}}$ & 5 &  7  &  8 &   {\bf{11}} &  4 \\
   $\rm{No\; Fault}$ &  {\bf{10}} &   9 &  7 &   4 &   5\\
\hline
\end{tabular}
\label{table:conf_50}
}

\vspace{-4mm}
\end{table}
The accuracy ($AC$), precision ($P$), and the false positive rate ($FP$) of the three algorithms are also evaluated from the confusion matrix results according to the following formulae \cite{confusion_2},
\vspace{-2mm}
\begin{eqnarray}
AC =\frac{\sum^5_{j=1}c_{jj}}{\sum ^5_{i=1}\sum^5_{j=1}c_{ij}},\;
P_j = \frac{c_{jj}}{\sum^5_{i=1}c_{ij}},\;
FP = \frac{\sum^4_{j=1} c_{5j}}{\sum^5_{j=1}c_{5j}},\notag
\end{eqnarray}
where $c_{ij},\; i,j=1,...,5$ denote the elements of the confusion matrix. {In Table \ref{table:conf_results}, the confusion matrix results according to the above metrics for the Tables \ref{table:conf_50}-\ref{table:conf_RPE} are provided. The results demonstrate that the accuracy of the fault diagnosis for the dual PE-based estimation algorithm outperforms RML method with $7.43\%$ and the false positive rate of $5.71\%$ less than RML method. The precision of the algorithm for all the system four health parameters is more than the ones from RML method. However, the Bayesian KS-based method indicates poor accuracy and high false alarm rate for the fault diagnosis of the system. Consequently, the PE-based method with $N=50$ \textit{outperforms} the other two methods \textit{significantly} in terms of \textit{higher accuracy, lower false positive rate, and higher precision} for all the four health parameters of the gas turbine engine.}
\vspace{-4mm}
\begin{table}[H]
\caption{{Confusion matrix Analysis results.}}
\small{
\label{table:conf_results}
\vspace{-5mm}
\begin{center}
\small{
\begin{tabular}{|c|c|c|c|c|c|c|c|}
\hline
Noise Level &$AC\%$& $FP\%$& $P_{\eta_{\rm{C}}}\%$& $P_{\dot m_{\rm{C}}}\%$ & $P_{\eta_{\rm{T}}}\%$& $P_{\dot m_{\rm{T}}}\%$   \\
\hline
\hline                  
  PE-based  ($50$ Part.) &86.29  &  5.71   & 93.94  &  93.75 & 77.78  &  74.36\\
    RML  ( $150$ Part.) &78.86   &11.43   &87.50   &72.97  &74.29  &72.22\\
    Bayesian KS-based  ($45$ Part.) & 25.95   &85.71   &25.00   &32.50   &23.68   &34.38 \\
\hline
\end{tabular}
}
\end{center}
\label{table:conf_results}
}
\end{table}
\vspace{-6mm}
 \section{Conclusion}\label{VI}
 \vspace{-1mm}
 In this paper, a novel dual estimation filtering scheme is proposed and developed based on particle filters (PF) to estimate a nonlinear stochastic system states and time variations in its parameters. The dual structure is based on the extension of the Bayesian parameter estimation framework. A dual structure is proposed for achieving simultaneous state and parameter estimation objectives. Performance results of the application of our method to a gas turbine engine under healthy and faulty scenarios are provided to demonstrate and illustrate the superior capability and performance of our scheme for a challenging fault diagnostic application {as compared to the well-known recursive maximum likelihood (RML) method based on particle filters and conventional Bayesian method for combined state and parameter estimation based on particle filters while the computational complexity of all the algorithms remains the same. On the other hand, the false alarm rates of our proposed dual algorithm is \textit{significantly lower} than the RML and conventional Bayesian methods. These \textit{two main characteristics} justify and substantiate the observation that our proposed algorithm is more suitable for the purpose of fault diagnosis of critical nonlinear systems that require lower fault detection times and false alarm rates}. Moreover, the estimation results accuracy in terms of the fault identification are also provided. The obtained results are demonstrated and validated by performing a confusion matrix analysis. The verification and validation of our results to real gas turbine engine is a topic of our future research.
 \vspace{-2mm}
\bibliographystyle{IEEEtr}
\bibliography{ref}
\appendix
{\bf{A: Proof of Proposition 1:}}
Let us consider the modified artificial evolution law by assuming that $A=I$ in equation \eqref{descent2}. Let us substitute $m^{(j)}_t$ from equation \eqref{descent3} where the superscript $(j)$ is omitted in order to define the modified artificial evolution law in a more general form that is also applicable to each single particle as
 \begin{eqnarray}\label{kernel_1}
 \tilde \theta_{t|t}= \hat\theta_{t-1|t-1}+P_t\psi _t\epsilon(t,\hat \theta_{t-1|t-1})+\zeta_t,
 \end{eqnarray}
where $P_t=\gamma_t R_t$. Now, let $V(.|y_{1:t})$ denote the variance of the stochastic process assuming that the observations up to time $t$ are available, and $C(. , .|y_{1:t})$ denotes the covariance of the two stochastic processes by assuming that the observations up to time $t$ are available. By taking into account the relationship between the variance of both sides of equation \eqref{kernel_1} when the covariance matrix is assumed to be non-singular and when the prediction error at time $t$ is uncorrelated with the parameter estimate at time $t$, given that $\hat\theta_{t-1|t-1}$ is independent of $P_t\Psi\epsilon_t$, therefore we get $ V(\tilde \theta_{t|t}|y_{1:t})=V(\hat \theta_{t-1|t-1}|y_{1:t})+P_{t}^{2}\Psi V_y\Psi^{\rm{T}}+W_{t}+2C(\hat \theta_{t-1|t-1},\zeta_{t}|y_{1:t})+2C(P_t\Psi\epsilon_t,\zeta_{t}|y_{1:t})$. Furthermore, since $P_t=\gamma_t R_t=\gamma_t \sqrt{{\rm{trace}}(\mathcal{E}_t\mathcal{E}_t^{\rm{T}})}$ is a scalar, one can write $ V(P_t\Psi\epsilon_t|y_{1:t})=(\mathbb E[P_t|y_{1:t}])^2V(\Psi\epsilon_t|y_{1:t})+(\mathbb E[\Psi\epsilon_t|y_{1:t}])^2V(P_t|y_{1:t})+V(P_t|y_{1:t})V(\Psi\epsilon_t|y_{1:t})= P_{t}^2\Psi V_t \Psi^{\rm{T}}$.

In order to ensure that there is no information loss (particularly in the case that $\theta_t$ is constant), one must have, $ V(\tilde \theta_{t|t}|y_{1:t})=V(\hat \theta_{t-1|t-1}|y_{1:t})=V_{\hat \theta_{t-1|t-1}}$, which implies that, $
 C(\hat \theta_{t-1},\zeta_{t}|Y_t)+C(P_t\Psi\epsilon_t,\zeta_{t}|Y_t) = -\frac{1}{2}W_{t}-\frac{1}{2}P_{t}^{2}\Psi V_t\Psi^{\rm{T}}$. Therefore, negative correlations are needed to remove the effects of unwanted information loss. In case of approximate joint normality of the stochastic process $(\hat\theta_{t-1|t-1},\zeta_{t}|Y_t)$ and $(P_t\Psi\epsilon_t,\zeta_{t}|Y_t)$, the conditional normal evolution is obtained as
{\small{
\begin{align}\label{shrinkage}
\begin{split}
 &p(\hat \theta_{t|t}|\hat \theta_{t-1|t-1})\thicksim
  \mathcal N(\hat \theta_{t|t}|A_{t}\tilde \theta_{t|t}+(I-A_{t})\hat\theta_{t-1|t-1},(I-A^{2}_{t})V_{\hat \theta_{t-1|t-1}}),
  \end{split}
 \end{align}}}
where the mean of this Gaussian distribution at each time step is found from equation \eqref{descent3}, when $\hat\theta_{t-1|t-1}$ is substituted by its modified version according to the shrinkage kernel. The shrinkage matrix $A_{t}$, is obtained as $ A_{t}= I - [\frac{1}{2}(W_{t}V_{\theta_{t-1|t-1}}^{-1}+P^{2}_{t}\Psi V_y\Psi^{\rm{T}}V_{\theta_{t-1|t-1}}^{-1})]$. Assuming that in the kernel shrinkage method, the variance of the evolution noise is interpreted as $W_t=(I-A^{2}_{t})V_{\hat \theta_{t-1|t-1}}$, therefore by replacing $W_t$ in the equation that was obtained for $A_t$ results in
\begin{eqnarray}\label{kernel_2}
A_{t}= I - \frac{1}{2}[(I-A_t^2)+P^{2}_{t}\Psi V_t\Psi^{\rm{T}}V_{\theta_{t-1|t-1}}^{-1}].
\end{eqnarray}

Let us assume that our main goal is to obtain and determine the shrinkage matrix $A_t$ as $A=aI$, therefore the matrix equation \eqref{kernel_2} can be written as
\begin{eqnarray}\label{kernel_3}
(a^2-2a+1)I-P^{2}_{t}\Psi V_t\Psi^{\rm{T}}V_{\theta_{t-1|t-1}}^{-1}=0.
\end{eqnarray}

We are interested in obtaining an upper bound for the shrinkage matrix that can be used for all time. Assuming that the last term in the right hand side of equation \eqref{kernel_2} has an upper bound given by $|P^{2}_{t}\Psi V_t\Psi^{\rm{T}}V_{\theta_{t-1|t-1}}^{-1}|\le P^{2}_{\rm{max}}\Psi V_y\Psi^{\rm{T}}V_{\hat \theta}^{-1}$, where $P_{\rm{max}}=\gamma_0 \sqrt{{\rm{trace}}(\mathcal{E}_{\rm{max}}\mathcal{E}_{\rm{max}}^{\rm{T}})}$ with $\gamma_0$ denoting the initial step size, therefore $\mathcal{E}_{\rm{max}}$ is considered as the maximum acceptable variance of the prediction error, $V_y$ is an upper bound of the measurement noise covariance, and $V_{\hat\theta}$ is the minimum bound of the parameter estimation covariance that is considered to be similar to the initial covariance of the parameters (before adding the evolution noise in time). Therefore, a bound on $aI$ and consequently $A$ can be obtained as $A=aI\le I - \sqrt{\frac{\sigma_{\rm {min}}(P^{2}_{\rm{max}}\Psi V_y\Psi^{\rm{T}}V_{\hat\theta}^{-1})}{{\sigma_{\rm {max}}(P^{2}_{\rm{max}}\Psi V_y\Psi^{\rm{T}}V_{\hat\theta}^{-1})}}}I$, where the normalization of the eigenvalue is performed to ensure that the associated fraction remains less than $1$. Let $a=1 - \sqrt{\frac{\sigma_{\rm {min}}(P^{2}_{\rm{max}}\Psi V_y\Psi^{\rm{T}}V_{\hat\theta}^{-1})}{{\sigma_{\rm {max}}(P^{2}_{\rm{max}}\Psi V_y\Psi^{\rm{T}}V_{\hat\theta}^{-1})}}}$, therefore, the shrinkage matrix becomes $A=aI$. The smoothing matrix corresponding to the normal distribution variance is now obtained from the shrinkage factor as $(1-a^{2})I$. This guarantees that the distribution \eqref{shrinkage} has a finite variance as $t\rightarrow \infty$ for both constant and time-varying parameter estimation cases. This completes the proof of the proposition.$\blacksquare$\\
{\bf{B: Proof of Theorem 1:}}  
The existence of the projection algorithm in the parameter estimation scheme ensures that $\tilde{\theta}^{(j)}_{t|t}$ remains inside $D_{\mathcal{N}}$. According to equation \eqref{posteriori_2}, the {\it{a posteriori}} estimate of the parameter at time $t$ is obtained from the resampled particles of the parameter estimate $\hat \theta^{(j)}_{t|t}$, as $\hat \theta_{t|t}=\frac{1}{N}\sum^{N}_{j=1}\hat \theta^{(j)}_{t|t}$, where $\hat \theta_{t|t}^{(j)}$ is selected from the $N$ particles of $\tilde \theta_{t|t}^{(j)}$ for which $\rho_{\nu_t}(y_t-h(\hat x_{t|t},\tilde \theta _{t|t}^{(j)}))$ yields higher probabilities. In order to avoid the discontinuity that is caused by resampling, in this procedure only the particles that are maintained from time $t-1$, i.e. $\hat \theta _{t-1|t-1}^{(j)}$ and are propagated to time $t$ as $\hat \theta_{t|t}^{(j)}$, are considered. However, the rest of the particles that are to be discarded in the resampling process will be replaced by the kept particles. Therefore, the results can be generalized to all the particles.

Consider equation \eqref{descent2} for generating $\tilde \theta^{(j)}_{t|t}$ and let us substitute $m_t^{(j)}$ from the PE-based update rule of \eqref{descent3}-\eqref{descent2} to obtain the following expression for the resampled particles $\hat \theta_{t|t}^{(j)}$, namely
\vspace{-4mm}
{\footnotesize{
\begin{eqnarray}\label{new_representation}
\begin{split}
\hat \theta_{t|t}^{(j)}=&A\hat \theta_{t-1|t-1}^{(j)}+(I-A)\frac{1}{N}\sum_{i=1}^N\hat \theta ^{(i)}_{t-1|t-1}   \\
&+A\gamma_t R_t^{(j)}\psi _t^{(j)}\epsilon(t,\hat \theta_{t-1|t-1}^{(j)})+ \sqrt{I-A^2}\zeta_t^{(j)},
\end{split}
\end{eqnarray}}}where $\sqrt{I-A^2}\zeta_t^{(j)}$ denotes the evolution noise by taking into account the kernel smoothing concept. By applying the sum operator to both sides of equation \eqref{new_representation} to construct $\hat \theta_{t|t}$ yields,
$\frac{1}{N}\sum^N_{j=1}\hat \theta_{t|t}^{(j)}=A\frac{1}{N}\sum^N_{j=1}\hat \theta_{t-1|t-1}^{(j)}+\frac{1}{N}\sum^N_{i=1}\frac{1}{N}\sum_{i=1}^N\hat \theta ^{(j)}_{t-1|t-1}-A\frac{1}{N}\sum^N_{i=1}\frac{1}{N}\sum_{i=1}^N\hat \theta ^{(j)}_{t-1|t-1}+A\frac{1}{N}\sum^N_{j=1}\gamma_t R_t^{(j)}\psi _t^{(j)}\epsilon(t,\hat \theta_{t-1|t-1}^{(j)})+\sqrt{I-A^2}\frac{1}{N}\sum^N_{j=1}\zeta_t^{(j)}=\frac{1}{N}\sum^N_{j=1}\hat \theta ^{(j)}_{t-1|t-1}+A\frac{1}{N}\sum^N_{j=1}\gamma_t R_t^{(j)}\psi _t^{(j)}\epsilon(t,\hat \theta_{t-1|t-1}^{(j)})$, which results in $\hat \theta_{t|t}=\hat \theta_{t-1|t-1}+A\frac{1}{N}\sum^N_{j=1}\gamma_t R_t^{(j)}\psi _t^{(j)}\epsilon(t,\hat \theta_{t-1|t-1}^{(j)})$. Assumptions A1 and A2 ensure that the regularity conditions are satisfied according to \cite{Walid}. Consequently, a differential equation associated with \eqref{descent3}-\eqref{descent2} can be obtained by considering that $\Delta \tau$ is a sufficiently small number and $t,\check{t}$ are specified such that $\sum_{k=t}^{\check{t}}\gamma_k=\Delta \tau$. Through a change of time-scales as $t\rightarrow \tau$ and $\check{t}\rightarrow \tau+\Delta \tau$, for a sufficiently small $\Delta \tau$, and by assuming that $\hat \theta_{t-1|t-1}=\check \theta, R^{(j)}_t=\check R^{(j)},\; A=aI$ is a constant matrix, the difference equation for $\hat \theta^{(j)}_{t|t}$ is now expressed as
\vspace{-2mm}
\begin{eqnarray}\label{difference_eq}
\hat \theta^{(j)}_{\check t}\approx \check \theta^{(j)}+ a \Delta \tau \check R^{(j)}  g(\check \theta^{(j)}),
\end{eqnarray}
where $g(\check \theta^{(j)})=\frac{1}{\Delta \tau}\sum_{k=t}^{t=\check t}\psi^{(j)}_t\epsilon(k,\hat \theta^{(j)}_{t-1|t-1})$. In the above derivation it is assumed that the $\hat \theta_{t|t}^{(j)}$ particle is kept after resampling (that is $\hat \theta^{(j)}_{t-1|t-1} \rightarrow \tilde \theta^{(j)}_{t|t} \rightarrow \hat \theta^{(j)}_{t|t})$. Consequently, considering Assumption A3, the differential equation associated with the evolution of each single particle is obtained as,
 \vspace{-1mm}
\begin{align}\label{d.e1}
\begin{split}
\frac{{\rm{d}}\theta_D^{(j)}}{{\rm{d}}\tau}=a R^{(j)}_D(\tau)g(\hat \theta^{(j)}_D(\tau))=-a R_D^{(j)}(\tau)[\frac{\rm{d}}{{\rm{d}}\hat \theta^{(j)}_D}\bar{J}(\hat\theta^{(j)}_D)]^{\rm{T}},
\end{split}
\end{align}
where the subscript $D$ is used to differentiate the solution of the differential equation \eqref{d.e1} from the solution of the difference equation \eqref{difference_eq}. Now, the required convergence analysis is reduced to investigating the properties of the deterministic continuous-time system \eqref{d.e1}.

Consider the positive definite function $L(\hat \theta^{(j)}_{t-1|t-1})=\mathbb{E}(\bar J(\hat \theta^{(j)}_{t-1|t-1}))=\frac{1}{N}\sum^N_{j=1}\bar J(\hat \theta^{(j)}_{t-1|t-1})$ that represents the expectation of a positive definite function through $N$ data points for $\hat \theta^{(j)}_{t-1|t-1}$. Our goal is to evaluate the derivative of this function along the trajectories of the system \eqref{d.e1}. According to Assumption A2, the second derivative of $Q(\epsilon(t,\hat \theta_{t-1|t-1}^{(j)}))$ is bounded, therefore the summation and derivative operations commute. According to Assumption A3 for $\hat \theta^{(j)}_{t-1|t-1}\in D_{\mathcal{N}}$, $\acute{\bar{J}}(\hat\theta^{(j)}_D(\tau))=\frac{\rm{d}}{\rm{d}\hat \theta^{(j)}_{t-1|t-1}}\bar{J}(\hat\theta^{(j)}_{t-1|t-1})|_{\hat\theta^{(j)}_{t-1|t-1}=\check \theta^{(j)}}=\bar {\mathbb{E}}(\frac{\rm{d}}{\rm{d}\hat\theta_{t-1|t-1}}Q(\epsilon(t,\hat\theta^{(j)}_{t-1|t-1}))$, exists and is approximated by $-g(\check \theta^{(j)})$. Therefore, let us define $V(\hat \theta^{(j)}_D)=\mathbb{E}(\bar J(\hat \theta_D^{(j)}(\tau))$, and given that $a>0$, and $R^{(j)}_D(\tau)$ is a positive scalar for $j=1,...,N$ (which represents the trace of a positive definite matrix at time $\tau$), one gets
\vspace{-2mm}
{\small{
\begin{eqnarray}
\begin{split}
&\frac{\rm{d}}{\rm{d}\tau}V(\hat \theta^{(j)}_D)=\mathbb{E}(\frac{\rm{d}}{\rm{d}\tau}\bar J(\hat \theta_D^{(j)}(\tau))=\frac{1}{N}\sum_{j=1}^N\acute{\bar J}(\hat \theta_D^{(j)})\frac{\rm{d}}{\rm{d}\tau}\hat \theta_D^{(j)}(\tau)\\
&=\frac{-a}{N}\sum_{j=1}^N[g(\hat \theta_D^{(j)}(\tau))]R^{(j)}_D(\tau)[g(\hat \theta_D^{(j)})(\tau)]^{\rm{T}}\le 0, \label{single}
\end{split}
\vspace{-4mm}
\end{eqnarray} }}where the equality is obtained only for $\hat\theta_D(\tau)\in D_\mathcal{C}$. Therefore, as $t\rightarrow \infty$ either $\hat \theta^{(j)}_{t-1|t-1}$ and consequently, $\pi^N_{\theta_{t-1|t-1}}$ w.p.1 tends to $D_{\mathcal{C}}$ or to the boundary of $D_{\mathcal{N}}$, where w.p.1 is with respect to the random variables related to the parameter estimate particles. It should be noted that for particles that have been replaced in the resampling this equality is valid since they are replaced by particles that have satisfied \eqref{single}. This completes the proof of the theorem.$\blacksquare$\\

\bigskip
{\bf{{C: Complexity Analysis Results:}}}  
{In the following Tables \ref{table:state}-\ref{table:baysian} the equivalent flop (EF) complexity associated with the \textit{three}  methods namely, (a) our proposed dual estimation scheme, (b) the conventional Bayesian method for state and parameter estimation \cite{Parameter} (where Regularized particle filter structure is utilized to implement the filter), and (c) the recursive maximum likelihood (RML) method according to simultaneous perturbation stochastic approximation (SPSA) algorithm \cite{poyi_gradientfree} are presented, respectively.}
\begin{table*}
\caption{{The Equivalent Complexity for the state estimation step in the Dual Structure.}}
\centering
\label{table:state}
   \scalebox{0.8}{
   \centering
\begin{tabular}{|c|c|c|c|c|}
\hline
\multirow{1}{*}{Instruction} & \multicolumn{1}{c|}{Mult.} & \multicolumn{1}{c|}{Add}& \multicolumn{1}{c|}{Func. Eval.} & \multicolumn{1}{c|}{Other}\\ 
\hline
$[U_1,T_1]={\rm{schur}}(\Sigma_{\hat x_{t-1|t}})$  &     $- $&     $- $&     $- $&   $10n_x^3 $  \\
$R_1={\rm{randn}}(n_x,N)$  &     $- $&     $- $&     $- $&   $N n_x c_1 $  \\
$\omega^{(i)}_{t}=(U_1\sqrt{T_1})R_1$&$n_x^3+Nn_x^2 $&$(n_x-1)n_x^2+N(n_x-1)n_x $&$- $& $n_x^2 $  \\
$\hat x_{t|t-1}^{(i)}=f_t(\hat x_{t-1|t-1}^{(i)},\hat{\theta}_{t-1|t-1}^{\rm{T}}\lambda(\hat x_{t|t-1}^{(i)}),\omega^{(i)}_{t})$&$-$&$-$&$Nn_x$& $-$  \\
$\hat y^{(i)}_{t|t-1}=h_t(\hat x_{t|t-1}^{(i)},\hat{\theta}_{t-1|t-1}^{\rm{T}}\lambda(x_{t|t-1}^{(i)}))$&$-$&$-$&$Nn_y$& $-$  \\
$\Sigma_{\hat x_{t|t-1}}=\frac{1}{N-1}\sum_{i=1}^N (\bar x_{t|t-1}^{(i)}-\hat x_{t|t-1})(\bar x_{t|t-1}^{(i)}-\hat x_{t|t-1})^{\rm{T}}$&$Nn_x^2$&$2Nn_x$&$-$& $-$  \\
{\rm{Reguralization and resampling to find weights}} ${w}_{x_{t}}^{(i)}$ and $\hat x^{(i)}_{t|t}$&$-$&$-$&$-$& $Nn_xc_2+Nn_xc_3$  \\
$\hat x_{t|t}=\frac{1}{N}\sum_{i=1}^{N}\hat x_{t|t}^{(i)}$&$n_x$&$Nn_x$&$-$& $-$  \\
 \hline
Total &$n_x^3+2Nn_x^2+n_x$&$n_x^3+(N-1)n_x^2+2Nn_x$&$N(n_x+n_y)$& $10n_x^3+n_x^2$  \\
 $\;$ &$\;$&$\;$&$\;$& $Nn_x(c_1+c_2+c_3)$  \\
 \hline
\end{tabular}
}
\end{table*}
\begin{table*}
\caption{{The Equivalent Complexity for the parameter estimation step in the Dual Structure.}}
\centering
\label{table:parameter}
   \scalebox{0.8}{
   \centering
\begin{tabular}{|c|c|c|c|c|}
\hline
\multirow{1}{*}{Instruction} & \multicolumn{1}{c|}{Mult.} & \multicolumn{1}{c|}{Add}& \multicolumn{1}{c|}{Func. Eval.} & \multicolumn{1}{c|}{Other}\\ 
\hline
$\bar y_{t|t-1}^{(j)}=h_t(\hat x_{t|t},\theta ^{(j)^{\rm{T}}}_{t-1|t-1}\lambda(\hat x_{t|t}))$  &   $- $&  $- $&  $Nn_y $&   $- $  \\
$\Sigma_{\theta}=(I-A^2)\Sigma_{\hat \theta_{t-1|t-1}}$  &   $n_{\theta}^3 $&  $(n_{\theta}-1)n_{\theta}^2+n_{\theta}^2 $&  $- $&   $- $  \\
$\epsilon^{(j)}_{t}=y_{t}-\bar{y}^{(j)}_{t|t-1}$  &     $- $&     $Nn_y $&     $- $&   $- $  \\
$\psi_{t}^{(j)}=\frac{dh}{d\theta}|_{\hat x_{t|t},\theta^{(j)}_{t-1|t-1}}$  &   $- $&  $- $&  $n_yn_{\theta} $&   $- $  \\
$P_{t}^{(j)}=\gamma_{t} (\sqrt{{\rm{trace}}(\epsilon^{(j)}_{t}\epsilon_{t}^{(j)^{\rm{T}}}})$  &  $N+Nn_y$&     $N(n_y-1)+Nn_y $&     $- $&   $- $  \\
$[U_2,L_2]={\rm{schur}}(\Sigma_{\theta})$ &     $-$&     $- $&     $- $&   $10n_{\theta}^3 $  \\
$R_2={\rm{randn}}(n_{\theta},N))$ &     $-$&     $- $&   $- $&   $Nn_{\theta}c_1 $  \\
$\zeta_{t}^{(j)}=(U_2 \sqrt{L_2})R_2$ &$n_{\theta}^3+Nn_{\theta}^2$&$(n_{\theta}-1)n_{\theta}^2+N(n_{\theta}-1)n_{\theta} $&   $n_{\theta}^2 $&   $- $  \\
$m_{t}^{(j)}=\hat{\theta}_{t-1|t-1}^{(j)}+P_{t}^{(j)}\psi_{t}^{(j)}\epsilon^{(j)}_{t}$& $N(n_y n_{\theta}+n_{\theta})$ &$N(n_y-1)n_{\theta}+Nn_{\theta}$& $- $&   $- $ \\
$\tilde {\theta}^{(j)}_{t|t}=Am^{(j)}_{t}+(I-A)\frac{1}{N}\sum^N_{j=1}\hat \theta^{(j)}_{t-1|t-1}+\zeta_{t}^{(j)}$&$Nn_{\theta}^2+n_{\theta}$&$Nn_{\theta}^2+2Nn_{\theta}+Nn_{\theta}+n_{\theta}^2$&$-$& $-$  \\
$\bar y_{t|t}=h_t(\hat x_{t|t},\tilde{\theta}_{t|t}^{(j)^{\rm{T}}}\lambda(\hat x_{t|t}))$&$-$&$-$&$Nn_y$& $-$  \\
{\rm{Resampling to find weights}}, ${w}_{\theta_{t}}^{(j)}$, {\rm{and}} $\hat \theta^{(j)}_{t|t}$&$-$&$-$&$-$& $Nn_{\theta}c_2$  \\
$\hat \theta_{t|t}=\frac{1}{N}\sum_{j=1}^{N}\hat \theta_{t|t}^{(j)}$&$n_\theta$&$Nn_\theta$&$-$& $-$  \\
$\Sigma_{\hat \theta_{t|t}}=\frac{1}{N-1}\sum_{i=1}^N (\hat \theta_{t|t}^{(j)}-\hat \theta_{t|t})(\hat \theta_{t|t}^{(j)}-\hat \theta_{t|t})^{\rm{T}}$&$Nn_{\theta}^2$&$2Nn_{\theta}$&$-$& $-$   \\
\hline
Total &$2n_{\theta}^3+3Nn_{\theta}^2$&$2n_{\theta}^3+2Nn_{\theta}^2$&$n_{\theta}^2+2Nn_y$& $10n_{\theta}^3+Nn_{\theta}c_1+Nn_{\theta}c_2$  \\
$\quad$ &$+(N+2)n_{\theta}+Nn_{\theta}n_y$&$+5Nn_{\theta}+3Nn_y-N$&$+n_yn_{\theta}$&  \\
$\quad$ &$+N(n_y+1)$ &$+Nn_\theta n_y$ &$\quad$ &\\
 \hline
\end{tabular}
}
\end{table*}
\begin{table*}
\caption{{The Equivalent Complexity for the Bayesian augmented state and parameter estimation scheme \cite{Parameter}.}}
\centering
\label{table:baysian}
   \scalebox{0.8}{
   \centering
\begin{tabular}{|c|c|c|c|c|}
\hline
\multirow{1}{*}{Instruction} & \multicolumn{1}{c|}{Mult.} & \multicolumn{1}{c|}{Add}& \multicolumn{1}{c|}{Func. Eval.} & \multicolumn{1}{c|}{Other}\\ 
\hline
$[U_1,T_1]={\rm{schur}}(\Sigma_{x,\theta})$  &     $- $&     $- $&     $- $&   $10(n_x+n_\theta)^3 $  \\
$R_1={\rm{randn}}(n_x+n_\theta,N)$  &     $- $&     $- $&     $- $&   $N (n_x+n_\theta) c_1 $  \\
$\omega^{(i)}_{t}=(U_1\sqrt{T_1})R_1$&$(n_x+n_\theta)^3+N(n_x+n_\theta)^2 $&$(n_x+n_\theta-1)(n_x+n_\theta)^2 $&$- $& $(n_x+n_\theta)^2 $  \\
$\;$&$\; $&$+N(n_x+n_\theta-1)(n_x+n_\theta)$&$\; $& $\; $  \\
$\omega^{(i)}_{x_{t}}=\omega^{(i)}_{t}(1:n_x) $&$-$&$- $& $- $ & $- $ \\
$\omega_{\theta_{t}}=(I-A^2)\omega^{(i)}_{t}(n_x+1:n_x+n_\theta)$  &   $n_{\theta}^3 $&  $(n_{\theta}-1)n_{\theta}^2+n_{\theta}^2 $&  $- $&   $- $  \\
\mbox{state/parameter augmentation}: $[\hat x_{t|t-1}^{(i)};\hat \theta_{t|t-1}^{(i)}]=$&$\;$&$\;$&$\;$& $\;$  \\
$[f_t(\hat x_{t|t-1}^{(i)},\hat{\theta}^{(i)^{\rm{T}}}_{t|t-1}\lambda(\hat x_{t|t-1}^{(i)}),\omega^{(i)}_{x_{t}});\hat \theta^{(i)}_{t|t-1}]$&$-$&$-$&$N(n_x+n_\theta)$& $-$ \\
$\hat y^{(i)}_{t|t-1}=h_t(\hat x_{t|t-1}^{(i)},\hat{\theta}^{(i)^{\rm{T}}}_{t|t-1}\lambda(\hat x_{t|t-1}^{(i)}))$&$-$&$-$&$Nn_y$& $-$ \\
$\Sigma_{x,\theta}=\frac{1}{N-1}\sum_{i=1}^N([\bar x_{t|t-1}^{(i)};\bar \theta_{t|t-1}^{(i)}]-[\hat x_{t|t-1};\hat \theta_{t|t-1}])$     $\;$&     $\; $&     $\; $&   $\; $ &   $\; $  \\
$\times([\bar x_{t|t-1}^{(i)};\bar \theta_{t|t-1}^{(i)}]-[\hat x_{t|t-1};\hat \theta_{t|t-1}])^{\rm{T}}$&$N(n_x+n_\theta)^2$&$2N(n_x+n_\theta)$&$-$& $-$  \\
$\;$  &     $\; $&     $\; $&     $\; $&   $\; $  \\
{\rm{Reguralization to find weights and resampling}} &$\;$&$\;$&$\;$& $\;$  \\
{\rm{to find}}, $\hat x^{(i)}_{t|t}$ and $\hat \theta^{(i)}_{t|t}$ &$-$&$-$&$-$& $N(n_x+n_\theta)(c_3+c_2)$  \\
$[\hat x_{t|t};\hat \theta_{t|t}]=[\frac{1}{N}\sum_{i=1}^{N}\hat x_{t|t}^{(i)};\frac{1}{N}\sum_{i=1}^{N}\hat \theta_{t|t}^{(i)}]$&$n_x+n_\theta$&$N(n_x+n_\theta)$&$-$& $-$  \\
 \hline
Total &$2n_\theta^3+n_x^3+n_x^2(2N+3n_\theta)$&$n_x^3+2n_\theta^3+n_x^2(3n_\theta-1+N)$&$N(n_x+n_\theta)$& $10n_x^3+10n_\theta^3$  \\
$\;$ &$+n_\theta^2(3n_x+2N)+4N n_\theta n_x$&$+n_\theta^2(N-1+3n_x)$&$+Nn_y$& $n_x^2(30n_\theta+1)$  \\
 $\;$ &$+n_x+n_\theta$&$+n_xn_\theta(2N-2)$&$\;$& $+n_\theta^2(30n_x+1)$  \\
  $\;$ &$\;$&$\;$&$\;$& $+n_x(Nc_1+2n_\theta+Nc_3)$  \\
    $\;$ &$\;$&$\;$&$\;$& $+n_\theta(Nc_1+Nc_3)$  \\
 \hline
\end{tabular}
}
\end{table*}
\begin{table*}
\caption{{The Equivalent Complexity for the Recursive Maximum Likelihood (RML)  Parameter scheme based on Particle Filters Using SPSA \cite{poyi_gradientfree}.}}
\centering
\label{table:rml}
   \scalebox{0.8}{
   \centering
\begin{tabular}{|c|c|c|c|c|}
\hline
\multirow{1}{*}{Instruction} & \multicolumn{1}{c|}{Mult.} & \multicolumn{1}{c|}{Add}& \multicolumn{1}{c|}{Func. Eval.} & \multicolumn{1}{c|}{Other}\\ 
\hline
$[U_1,T_1]={\rm{schur}}(\Sigma_x)$  &     $- $&     $- $&     $- $&   $10n_x^3 $  \\
$R_1={\rm{randn}}(n_x,N)$  &     $- $&     $- $&     $- $&   $N n_x c_1 $  \\
$\omega^{(i)}_{t}=(U_1\sqrt{T_1})R_1$&$n_x^3+Nn_x^2 $&$(n_x-1)n_x^2+N(n_x-1)n_x $&$- $& $n_x^2 $  \\
\mbox{Generate random perturbation vector} $\Delta _t$  &     $- $&     $- $&     $- $&   $n_{\theta}c_1 $  \\
\mbox{for} $i=1,...,N$ \mbox{sample}: &     $- $&     $- $&     $- $&   $- $  \\
$\tilde{x}_{t,+}^{(i)}=f_t(\hat{x}_{t-1|t-1}^{(i)},(\hat{\theta} _{t-1|t-1}+c_t\Delta_t)^{\rm{T}}\lambda (\hat{x}_{t-1|t-1}^{(i)}),\omega_t^{(i)})$  &     $- $&     $- $&     $- $&   $N n_x $  \\
$\tilde{x}_{t,-}^{(i)}=f_t(\hat{x}_{t-1|t-1}^{(i)},(\hat{\theta} _{t-1|t-1}-c_t\Delta_t)^{\rm{T}}\lambda (\hat{x}_{t-1|t-1}^{(i)}),\omega_t^{(i)})$  &     $- $&     $- $&     $- $&   $N n_x $  \\
\mbox{Evaluate}: &     $- $&     $- $&     $- $&   $N (n_x+n_\theta) c_1 $  \\
$a_{\theta}(y_t,\tilde{x}_{t,+}^{(i)},\hat{x}^{(i)}_{t-1|t-1})$, \mbox{and} $a_{\theta}(y_t,\tilde{x}_{t,-}^{(i)},\hat{x}^{(i)}_{t-1|t-1})$  &     $- $&     $- $&     $- $&   $2N n_\theta$  \\
\mbox{Evaluate}: &     $- $&     $- $&     $- $&   $- $  \\
$\widehat{\nabla J}_{t,\mu}(\hat{\theta} _{t-1|t-1})=\frac{\hat{J}_t(\hat{\theta}_{t-1|t-1}+c_t\Delta _t)-\hat{J}_t (\hat{\theta}_{t-1|t-1}-c_t\Delta _t)}{2c_t\Delta_{t,\mu}}$  &     $n_{\theta}+1 $&     $2n_{\theta}-1 $&     $- $&   $- $  \\
\mbox{where}: &     $- $&     $- $&     $- $&   $- $  \\
$\widehat{\nabla J}_{t,\mu}(\hat{\theta} _{t-1|t-1}\pm c_t\Delta _t)=log \{ \frac{1}{N}\sum ^{N}_{i=1}a_{\theta}(y_t,\tilde{x}_{t,\pm}^{(i)},\hat{x}^{(i)}_{t-1|t-1})\}$  &     $2n_{\theta} $&     $2Nn_{\theta} $&     $2n_{\theta} $&   $- $  \\
\mbox{Parameter Update}: &     $- $&     $- $&     $- $&   $-$  \\
$\hat {\theta} _{t|t}=\hat{\theta}_{t-1|t-1}+\gamma _t \widehat{\nabla J}_t(\hat{\theta} _{t-1|t-1}),$  &     $n^2_{\theta} $&     $n_{\theta}+(n_{\theta} -1)n_{\theta} $&     $- $&   $- $  \\
$\widehat{\nabla J}_t(\hat{\theta}_{t-1|t-1})=[\widehat{\nabla J}_{t,1}(\widehat{\theta}_{t-1|t-1}),...,\widehat{\nabla J}_{t,n_{\theta}}(\hat{\theta}_{t-1|t-1})]$  &     $- $&     $- $&     $- $&   $- $  \\
\mbox{for} $i=1,...,N$ \mbox{sample}: &     $- $&     $- $&     $- $&   $- $  \\
$\tilde{x}_{t}^{(i)}=f_t(\hat{x}_{t-1|t-1}^{(i)},\hat{\theta} _{t|t}^{\rm{T}}\lambda (\hat{x}_{t-1|t-1}^{(i)}),\omega_t^{(i)})$  &     $- $&     $- $&     $Nn_x $&   $- $  \\
\mbox{Regularization to evaluate weights and resampling}  &     $- $&     $- $&     $- $&   $N n_x(c_2+c_3) $  \\
 \hline
Total &$n_x^3+Nn_x^2+n^2{_\theta}$&$n_x^3+(N-1)n_x^2-Nn_x$&$Nn_x+2n_\theta$& $10n_x^3+n_x^2$  \\
$\;$ &$+3n_\theta + 1$&$+n_{\theta}^2+(2N+2)n_{\theta}-1$& & $+n_x(2Nc_1+2N+Nc_2)$  \\
 $\;$ & & &$\;$& $+n_\theta(c_1+Nc_1+2N)$  \\
 \hline
\end{tabular}
}
\end{table*}

\end{document}